\documentclass[twocolumn]{autart}    % Enable this line and disable the 
\usepackage{graphicx}          % Include this line if your 
\usepackage{amsmath}    		% Extra math definitions
\usepackage{graphics} 		% PostScript figures
\usepackage{setspace}		% 1.5 spacing
\usepackage{longtable}          % Tables spanning pages
\usepackage{verbatim}
\usepackage{fancyhdr}      % (Ali) To change the heading in the appendix
\usepackage{bm}
\usepackage[noadjust]{cite}

\usepackage{color}
\usepackage{pdfpages}
\usepackage{epsfig} % for postscript graphics files
\usepackage{mathptmx} % assumes new font selection scheme installed
\usepackage{times} % assumes new font selection scheme installed
\usepackage{amsmath} % assumes amsmath package installed
\usepackage{amssymb}  % assumes amsmath package installed
\usepackage{amsfonts}
\usepackage{tikz,tabstackengine,amsmath}
\usetikzlibrary{shapes,shadows,arrows}
\usetikzlibrary{shapes,shapes.geometric,arrows.meta,fit,calc,positioning,automata}
\usepackage{pgfplots}
\usepgfplotslibrary{groupplots}
\usepackage{subfig}
\usepackage{mathtools}
\usepackage{adjustbox}
\usepackage{xcolor}
\usepackage{color}
\usepackage [english]{babel}
\usepackage [autostyle, english = american]{csquotes}
\usepackage{todonotes}
\usepackage[utf8]{inputenc}

\usepackage{enumitem}
\newtheorem{remark}{Remark}

\newcommand{\cal}{\mathcal}
\newcommand{\mb}{\mathbb}
\newcommand{\mc}{\mathcal}

\newcommand{\hf}{\tfrac{1}{2}}

\newcommand\bovermat[2]{%
  \makebox[0pt][l]{$\smash{\overbrace{\phantom{%
    \begin{matrix}#2\end{matrix}}}^{\text{#1}}}$}#2}

\definecolor{myGrn}{rgb}{0,0.5,0}

\usepackage[normalem]{ulem}

\newcommand{\mf}[1]{{{\mbox{\Large \boldsymbol{#1}}}}}
\newcommand{\mfs}[1]{{{\mbox{\large \boldsymbol{#1}}}}}
\begin{document}

\begin{frontmatter}

\title{A Class of Hybrid LQG Mean Field Games with State-Invariant Switching and Stopping Strategies\thanksref{footnoteinfo}} % Title, preferably not more 
%\thanks[footnoteinfo]{This paper was not presented at any IFAC  meeting. Corresponding author M.~T.~Cicero. Tel. +XXXIX-VI-mmmxxi.  Fax +XXXIX-VI-mmmxxv.}

\thanks[footnoteinfo]{This work was supported in part by the Fonds de Recherche du Qu\'ebec--Nature et Technologies (funding reference number 258061) and in part by the Natural Sciences and Engineering Research Council, Canada. Corresponding author D.~Firoozi.}
%This work was supported in part by the Fonds de Recherche du Qu\'ebec--Nature et Technologies and in part by the Natural Sciences and Engineering Research Council, Canada.}

\author[Paestum]{Dena Firoozi}\ead{dena.firoozi@hec.ca},    % Add the 
\author[Rome]{Ali Pakniyat}\ead{apakniyat@ua.edu},               % e-mail address 
\author[Baiae]{Peter E Caines}\ead{peterc@cim.mcgill.ca}  % (ead) as shown

\address[Paestum]{Department
of Decision Sciences, HEC Montr\'eal, Montreal,
QC, Canada}  % Please supply                                              
\address[Rome]{Department of Mechanical Engineering, University of Alabama, Tuscaloosa, AL, USA}             % full addresses
\address[Baiae]{Centre for Intelligent Machines (CIM) and the Department of Electrical and Computer Engineering (ECE), McGill University, Montreal, QC, Canada}        % here.

                             % keyword list or with the 
                                          % help of the Automatica 
                                          % keyword wizard

\begin{abstract}                          % Abstract of not more than 200 words.
A novel framework is presented that combines Mean Field Game (MFG) theory and Hybrid Optimal Control (HOC) theory to obtain a unique $\epsilon$-Nash equilibrium for a non-cooperative game with switching and stopping times. We consider the case where there exists one major agent with a significant influence on the system together with a large number of minor agents constituting two subpopulations, each agent with individually asymptotically negligible effect on the whole system. Each agent has stochastic linear dynamics with quadratic costs, and the agents are coupled in their dynamics and costs by the average state of minor agents (i.e. the empirical mean field). 
 It is shown that for a class of Hybrid LQG MFGs, the optimal switching and stopping times are state-invariant and only depend on the dynamical parameters of each agent. Accordingly, a hybrid systems formulation of the game is presented 
%The hybrid feature enters 
via the indexing by discrete events: (i) the switching of the major agent between alternative dynamics or (ii) the termination of the agents' trajectories in one or both of the subpopulations of minor agents. Optimal switchings and stopping time strategies together with best response control actions for, respectively, the major agent and all minor agents are established with respect to their individual cost criteria by an application of Hybrid LQG MFG theory.
\end{abstract}

\end{frontmatter}

\section{Introduction}
Mean Field Game (MFG) theory studies the existence of approximate Nash equilibria and the corresponding individual strategies for stochastic dynamical systems in games involving a large number of agents. Basically, the theory exploits the relationship between the large finite and the corresponding infinite limit population problems. The equilibria are termed $\epsilon$-Nash equilibria and are generated by the local, limited information control actions of each agent in the population. The control actions constitute the best response of each agent with respect to the behaviour of the mass of agents. Moreover, the approximation error, induced by using the MFG solution, converges to zero as the population size tends to infinity.

The analysis of this set of problems originated in \cite{HuangCDC2003a,HuangCIS2006,HuangTAC2007}, and independently in \cite{Lasry2006a,Lasry2006b,Lasry2007}. Many extensions and generalizations of MFGs exist, principally the probabilistic formulation \cite{CarmonaDelarueBook2018}, the master equation approach \cite{CardaliaguetMasterEqBook2019} and mean field type control theory \cite{BensoussanBook2013}. In \cite{Huang2010, Huang2012} the authors analyse and solve the completely observed (CO) linear quadratic Gaussian (LQG) systems case where there is a major agent (i.e. non-asymptotically vanishing as the population size goes to infinity)  together with a population of minor  agents (i.e. individually asymptotically negligible). The existence of  closed-loop $\epsilon$-Nash equilibria is established together with the individual agents' control laws that yield the equilibria \cite{Huang2012}. A convex analysis method is utilized in \cite{FCJ-Convex2018} to retrieve the solutions of \cite{Huang2010}, where no assumption is imposed on the evolution of the mean field a priori. %For such systems, among others, \cite{HuangMA2020} presents a multi-scale analysis and the notion of asymptotic solvability, and \cite{Kordonis2015} considers the case with a random number of minor agents.
The CO MM nonlinear (NL) MFG problem is treated in \cite{NourianSiam2013}. This framework is further extended in \cite{KizilkaleTAC2016, SenSiam2016, FirooziCainesTAC2020, FirooziISDG2017, ThesisDena2019,FirooziCainesCDC2019} for partially observed MFG theory for nonlinear and linear quadratic systems. Using the probabilistic approach to MFGs, \cite{CarmonaZhu2016,CarmonaWang2017} establish the existence of open-loop and closed-loop $\epsilon$-Nash equilibria for a general MM MFG and provide explicit solutions for an LQG case.  The works \cite{LasryLions2018, CardaliaguetCirant2018} characterize the Nash equilibrium for a general MFG system with one major agent and an infinite number of minor agents via the MFG  Master Equations. It is to be noted that for the LQG case it has been, respectively, demonstrated in \cite{HuangCIS2020} and \cite{Firoozi2020} that the (Markovian) closed-loop solutions to LQG MM MFGs obtained through the master equation and the probabilistic approaches are identical to the original LQG MM MFG solutions of  \cite{Huang2010}. (Another line of research characterizes a Stackelberg equilibrium between the major agent and the minor agents, see e.g. \cite{BensoussanSICON2017,BasarMoon2018}.)

MFGs have found numerous applications in engineering problems such as cellular network optimization \cite{aziz2016mean} and coordination of loads in smart grids \cite{KIZILKALE2019} (see \cite{Tembine2017} for a set of interesting applications), and in particular in mathematical finance and economics for characterizing equilibrium price and market equilibria (see \cite{shrivats2020mean,FirooziISDG2017,Gomes2016, CarmonaRev2020} and the references therein) -- to name a few. 

In several situations in stochastic dynamic games, such as in financial markets \cite{FirooziPakniyatCainesCDC2017}, agents wish to find the best time at which to enter or exit a given strategy. In order to determine the optimal stopping time strategies together with best response policies for the agents one is required to invoke the necessary optimality conditions of stochastic hybrid optimal control theory \cite{APPEC2017IFAC, APPECCDC2016, AghayevaAbushov, BensoussanMenaldiStochastic}. These optimality conditions are an extension of deterministic hybrid optimal control theory \cite{BensoussanMenaldi, SussmannNonSmooth, GaravelloPiccoli, ShaikhPEC, FarzinPECSIAM, APPEC2017TAC, APPECTAC2020} for systems interacting with stochastic diffusions. In \cite{APPECCDC2016}, in particular, the Stochastic Hybrid Minimum Principle (SHMP) is established for a general class of stochastic hybrid systems with both autonomous and controlled switchings and jumps possibly accompanied by dimension changes. Given the computational difficulty in solving general nonlinear forward-backward stochastic differential equations (FB-SDE) and the associated boundary conditions via the SHMP, a class of linear quadratic Gaussian (LQG) HOC problems is presented in \cite{APPEC2017IFAC} for which the corresponding Riccati equations are independent of the realizations of the stochastic diffusion terms.

%The first combination of Mean Field Game (MFG) theory and Hybrid Optimal Control (HOC) theory appeared in \cite{FirooziPakniyatCainesCDC2017} in a non-cooperative game formulation of the electronic markets where high frequency traders (HFTs) may leave the market before a given final time. The best stopping time policies for the traders are further shown to yield a closed-loop \mbox{$\epsilon$-Nash} equilibrium for the market. In this paper, we extend these results and develop a hybrid MFG framework for a class of LQG MFGs with a hybrid major agent and multiple stopping subpopulations of minor agents whose optimal switching and stopping times are state-invariant; specifically a major agent is permitted to freely switch between different dynamics and the two subpopulations of minor agents are provided with the option to stop at some optimal time. Each agent has stochastic linear dynamics with quadratic costs, and the agents are coupled in their dynamics and costs by the average state of minor agents (the empirical mean field). Since the governing stochastic differential equations for the system change with the switching of the major agent or the cessation of one or both subpopulations of minor agents, a hybrid systems formulation of the problem is presented which employs an indexing of these modes by discrete events.

 The first combination of Mean Field Game (MFG) theory and Hybrid Optimal Control (HOC) theory appeared in the predecessor to the current paper, \cite{FirooziPakniyatCainesCDC2017}; in that analysis a non-cooperative game formulation of electronic markets was presented, where high frequency traders (HFTs) may leave the market before a given final time. The best stopping time policies for the traders are further shown to yield a closed-loop \mbox{$\epsilon$-Nash} equilibrium for the market. 
 %\ali{I recommend having a shorter version here (while keeping the long version in our response letter). Thus, instead of "While [15] ...." we shall say something like: "This work expands upon the authors' early work by presenting the theory in the general setting of time-varying LQG mean field games with switching and stopping time strategies where switching costs and jump maps are permitted to be time-varying and where the major agent is permitted to switch its own dynamics (and, hence, the game's dynamics). Furthermore, this work provides detailed statements of theorems and their proofs, including the sufficient optimality conditions for hybrid LQG optimal control and stopping problems, as well as a Dynamic Programming (DP)-based determination of the optimal sequence of the discrete events accompanying the Stochastic Hybrid Minimum Principle (SHMP)-based determination of the best responses. Last but not least, a numerical example is provided in this article to illustrate the results.}
 The advances in this paper beyond the contributions in \cite{FirooziPakniyatCainesCDC2017} are as follows:
\begin{itemize}
    \item The system considered is a general time-varying LQG mean field game system with switching and stopping time strategies. As such, the time derivative of the switching cost weight matrix appears in the switching (stopping) equation (11) (eq. (15)).
    \item Both sufficient and necessary optimality conditions for a switching (stopping) to take place are provided in Theorem \ref{thm:HybridLQG} and Corollary \ref{thm:HybridLQGstop}.% together with complete proofs. 
    \item The proofs of Theorem \ref{thm:HybridLQG}, Corollary \ref{thm:HybridLQGstop}, and Theorem~\ref{thm:ENE_hybrid} ($\epsilon$-Nash property) are presented. 
    %\item  The $\epsilon$-Nash Theorem \ref{thm:ENE_hybrid} is given and proved.
    \item A numerical methodology is developed for solving the set of hybrid MFG equations in Subsection 4.5 and is implemented for an example in Section 5. 
    \item The major agent is provided with the option to switch to different dynamics which leads to more complex automata. More specifically, the system has 4 discrete states with an associated increase in the number of potential realizations compared to the case study in \cite{FirooziPakniyatCainesCDC2017} where there exist 3 discrete states. %with a smaller number of realizations in each case.  
 \end{itemize}

We observe that for general Hybrid LQG problems (including LQG stopping systems as the special case with switching to zero dynamics), the optimal switching (and stopping) times are filtration-adapted random variables and, hence, optimal inputs are not necessarily representable in a Riccati format. For such problems, nonlinear versions of hybrid mean field games may be formulated and solved, which is beyond the scope of the current paper and is the subject of future work. However, as discussed in Theorem \ref{thm:HybridLQG} (and Corollary \ref{thm:HybridLQGstop}), for certain classes of hybrid (and stopping) LQG systems, the optimality conditions of the SHMP yield state-invariant representations of the optimal switching (and stopping) times, which can be identified deterministically based upon the dynamical parameters of each agent. Hence, in the limiting MFG formulation of the problem all minor agents within the same subpopulation stop at the same time yielding a deterministic representation of the mean field. Subsequently, a hybrid formulation of the game is developed for which switching events correspond to (i) the switching of the major agent or (ii) the cessation of one or both subpopulations of minor agents. Hence, by developing and then utilizing a hybrid LQG MFG theory, optimal switching and stopping time strategies for, respectively, the major agent and all minor agents, together with their best response control actions which yield a unique $\epsilon$-Nash equilibrium are established.

A recent work \cite{Tankov-SICON-2020} studies the stopping of agents in the infinite-population MFG systems, where a relaxed solution approach is followed by looking for the occupation measure of agents instead of their stopping time. \cite{Kordonis2015} studies MFG systems where at each time instant a random number of agents enter and remain in the system for a specific time duration. The random entrance is described by a Markov chain and the problem is formulated as an LQG optimal control problem for Markov jump linear systems.

We note that the following terms  are used interchangeably throughout the paper: optimal and best response in the infinite-population case, quit and stop, control action and control input. 

 The paper organization is as follows: Section \ref{sec:HybridLQG} introduces single-agent hybrid LQG systems with state-invariant switching and stopping strategies. Subsequently, Section \ref{sec:ProblemFormulation} presents the class of hybrid LQG MFG problems under study. More specifically, this section is devoted to the two types of transitions that exist between the dynamics at the individual level, i.e. at a transition event the major agent switches dynamics (from one of the realizations of equation \eqref{MajorStandardDynamics_hybrid} to the other) or a minor agent stops (switches from dynamics \eqref{MinorStandardDynamics_hybrid} to zero dynamics). Section \ref{sec:InfinitePopulation} presents hybrid-MFG approach, where, at each discrete state, the major agent's state is extended by the corresponding mean field, and a generic minor agent's state is extended by the corresponding major agent's state and the mean field. At the mean field game level the dynamics governing the extended states are undergoing changes, in which case, the associated dynamics are presented in Subsections \ref{sec:hybDynCost} and \ref{sec:hybDynCost-minor}, the corresponding transitions in Subsection \ref{sec:majorJumpMapNswitchingCost} and Appendix \ref{sec:minorJumpMapNSwitchingCost}, and the best-response solutions in Subsections \ref{sec:BRHybCA-major} and \ref{sec:BRHybCA-minor}, respectively, for the major agent and a generic minor agent. Subsequently, the hybrid-MFG consistency equation and $\epsilon$-Nash property are presented in Subsection \ref{sec:ConsistencyCondition}. Next, Section \ref{sec:SimulationHybridMFG} depicts simulation results. Finally, Section \ref{sec:ConclusionHybridMFG} presents concluding remarks and future directions.

%\begin{figure}
%\begin{center}
%\includegraphics[height=4cm]{jcaesar.eps}    % The printed column  
%\caption{Gaius Julius Caesar, 100--44 B.C.}  % width is 8.4 cm.
%\label{fig1}                                 % Size the figures 
%\end{center}                                 % accordingly.
%\end{figure}

% OR

%\begin{figure}
%\begin{center}
%\epsfig{file=jcaesar,width=7cm}
%\caption{Gaius Julius Caesar, 100--44 B.C.}
%\label{fig1}
%\end{center}
%\end{figure}
\section{State-Invariant Optimal Switching and Stopping Strategies for Single-Agent Hybrid LQG Systems}
\label{sec:HybridLQG}
In this section single-agent hybrid LQG systems are presented. Then, a set of sufficient conditions (stemming from \cite{APPEC2017IFAC}) are derived under which the optimal switching and stopping times for such systems are state-invariant. While \cite{APPEC2017IFAC} identifies a class of single-agent hybrid LQG systems for which the optimal switching times obtained from the Hamiltonian continuity condition do not depend on the state or the initial condition, its primary focus is on the necessity of these conditions. The extension of those results developed in this section are broader and cover a larger class of hybrid LQG systems. In particular,
\begin{itemize}
    \item The class of hybrid LQG problems considered in Section 2 include \textit{multi-variate time-varying} LQG systems and \textit{switching costs} are now incorporated in the cost functional. As a result, the time derivative of the switching cost weight matrix appears in the optimality necessary conditions,
    \item A set of sufficient conditions are provided for the filtration-invariance of switchings,
    \item The case where an agent is permitted to stop is presented as a special case of controlled switching and the corresponding necessary and sufficiency conditions are established accordingly. 
\end{itemize}

The results are subsequently used in the formulation of a class of hybrid LQG MFGs in the rest of the paper.  

%The following exposition is an elaboration of the results of \cite{APPEC2017IFAC} that presents a set of sufficient conditions under which the optimal switching and stopping times for LQG systems are \mbox{${{\cal F}_t}$-independent} and state-invariant.% and, consequently, for them to be almost surely equal for all agents within a subpopulation.

Let $\left(\Omega,{\cal F},\{{\cal F}_{t}\}_{t \in [0,T]},P\right)$ be a probability space such that ${\cal F}_0$ contains the $P$-null sets, ${\cal F}_{T}={\cal F}$ for a fixed final time $T < \infty$, and let ${\cal F}_t$ be the natural filtration associated with the sigma-algebra generated by a Wiener process up to time $t$.

The (hybrid) state of a stochastic hybrid system is denoted by $h=(Q,x)$ where $Q \in \mathbb{Q}$ denotes the discrete state (component) taking values from $\mathbb{Q}$ with finite cardinality, and $x \in \mathbb{R}^{n_Q}$ denotes the continuous component of the hybrid state (shortly referred to as the continuous state). %The discrete state $Q$ evolves discretely in time (i.e., it changes over a countable set of (switching) time instances, and the continuous component $x$ evolves continuously in time (a consequence of which is that the dynamics of $x$ satisfies stochastic differential equations \eqref{LinearSDE}).
We introduce the \textit{counting index} $j \in \mathbb{Z}_{\geq 0}$ that indicates the number of switchings incurred within the interval $[t_0,t)$. Conversely, denoting by $t_{j}$ the $j^{\text{th}}$ switching instant, the expression $t \in \left[t_{j},t_{j+1}\right)$ indicates that the value of $Q$ has changed (switched) $j$ times by time $t$. In  this paper, all changes in the value of $Q$ are controlled switchings, i.e. every switching is a direct consequence of a control action. 

A \textit{hybrid input process} is a pair $\left(S_L,u(\cdot)\right) =: I_{L}\equiv I_{L}^{[0,T]}$ defined on $\left[0,T\right]$, $T<\infty$, where $S_L = \big(\left(t_{0},Q_{0}\right), \left(t_{1},Q_{1}\right), \cdots,$ $\left(t_{L},Q_{L}\right) \big)$, $L<\infty$, is a finite \textit{hybrid sequence of switching events} consisting of a strictly increasing sequence of ${\cal F}_t$-adapted times $t_j$, %i.e. $t_{0} < t_{1} < t_{2} < \ldots < t_{L} < T =: t{L+1}$ almost surely, 
and $u(\cdot) := \{u^{Q_0}(\cdot), u^{Q_1}(\cdot), \cdots, u^{Q_L}(\cdot)\}$ is an ${\cal F}_t$-adapted  continuous input process, where for every $t \in [t_{j},t_{j+1}]$, $j \in \{0,1,\cdots,L\}$ (equivalently denoted by $0 \leq j \leq L$ or $j=0,1,\cdots,L$), the continuous input $u^{Q_j}(t)$ is an ${\cal F}_t$-adapted, $\mathbb{R}^{m_{Q_{j}}}$ valued, random variable.

%Given a hybrid input process $I_{L}$, the 
The dynamics of the continuous state process are %considered  to be 
governed by linear It\^o differential equations of the form
\begin{multline}
dx^{Q_{j}}(t)=\left(A^{Q_{j}}\left(t\right) x^{Q_{j}}\left(t\right) +B^{Q_{j}}\left(t\right)u^{Q_{j}}\left(t\right)\right)dt \\+D^{Q_{j}}\left(t\right)dw(t),\quad t\in\left[t_{j},t_{j+1}\right),
\label{LinearSDE}
\end{multline}
where $Q_{j} \in \mathbb{Q}$, $x^{Q_{j}}\left(t\right) \in \mathbb{R}^{n_{Q_{j}}}$, $u^{Q_{j}}\left(t\right) \in \mathbb{R}^{m_{Q_{j}}}$, $w(t) \in \mathbb{R}$,  $A^{Q_{j}}\left(t\right) \in \mathbb{R}^{n_{Q_{j}} \times n_{Q_{j}}}$, $B^{Q_{j}}\left(t\right) \in \mathbb{R}^{n_{Q_{j}} \times m_{Q_{j}}}$, $D^{Q_{j}}\left(t\right) \in \mathbb{R}^{n_{Q_{j}}}$, $0 \leq j \leq L$, $t_{0}:=0$, $t_{L+1}:=T$. 
%Switching from a discrete state $Q_{j-1} = q \in \mathbb{Q}$ to another discrete state $Q_{j} = q^{\prime} \in \mathbb{Q}$ is considered to be a controlled switchings, that is the direct result of a discrete input $\sigma_{j} \in \Sigma$ at an arbitrary ${\cal F}_t$-adapted switching time $t_j^{\omega}$. 
Upon switching at a switching time $t_j$, the continuous component of the state is reinitialized according to a jump map provided as
\begin{equation}
x^{Q_j}\left(t_{j}\right)=\Psi_{\sigma_{j}}x^{Q_{j-1}}\left(t_{j}-\right) \equiv \Psi_{\sigma_{Q_{j-1},Q_{j}}} \, x^{Q_{j-1}}\left(t_{j}-\right).
\label{JumpMap}
\end{equation}
It is further assumed that
\begin{equation}
D^{Q_{j}}(t_j)=\Psi_{\sigma_{{Q_{j-1}Q_{j}}}} D^{Q_{j-1}}(t_j), \label{A1EquivalentDiffusion}
\end{equation}
\\[-4pt]
\noindent for all $1 \leq j \leq L$, which implies equivalent diffusion fields before and after switching events.

Given an initial condition $\left(Q(0),x^{Q_0}(0)\right) = \big(Q_0,x^{Q_0}_0\big)$, the cost associated with the hybrid input $I_{L}$ over the  time horizon $\left[0,T\right]$ is considered to be of the form
\begin{multline}
J\Big(t_0, \big(Q_0,x^{Q_0}_0\big), I_L\Big)= \frac{1}{2} \mathbb{E}\Bigg[ \Vert x^{Q_{L}}(T)\Vert_{\bar{P}^{Q_{L}}(T)}^{2}
\\[-1pt]
+\sum_{j=1}^{L}\Vert x^{Q_{j-1}}(t_{j}-)\Vert_{C_{\sigma_{j}}(t_j )}^{2}
\allowdisplaybreaks
\\
+ \sum_{i=0}^{L}\int_{t_{i}}^{t_{i+1}}\Big(\left\Vert x^{Q_{i}}\left(t\right)\right\Vert_{P^{Q_{i}}\left(t\right)}^{2}+\left\Vert u^{Q_{i}}\left(t\right)\right\Vert_{R^{Q_{i}}\left(t\right)}^{2}\Big)dt \Bigg],
\label{HybridCostLQG}
\end{multline}
where $0 \leq \left[\bar{P}^{Q_{L}}\left(t\right)\right]^{T}=\bar{P}^{Q_{L}}\left(t\right) \in \mathbb{R}^{n_{Q_{L}} \times n_{Q_{L}}}$, $0 \leq \left[C_{\sigma_{j}}\left(t\right)\right]^{T}=C_{\sigma_{j}}\left(t\right) \in \mathbb{R}^{n_{Q_{j-1}} \times n_{Q_{j-1}}}$, $0 \leq \left[{P^{Q_{i}}}\left(t\right)\right]^{T}=P^{Q_{i}} \left(t\right) \in \mathbb{R}^{n_{Q_{i}} \times n_{Q_{i}}}$, $0 < \left[R^{Q_{i}} \left(t\right)\right]^{T}=R^{Q_{i}}\left(t\right) \in \mathbb{R}^{m_{Q_{i}} \times m_{Q_{i}}}$. 
The associated stochastic hybrid optimal control problem is to find $\inf_{I_L} J\big(t_0, (Q_0,x^{Q_0}_0), I_L\big)$.

\begin{thm}\label{thm:HybridLQG}
{\small \textbf{(Sufficient Conditions for ${\cal F}_t$-invariance of Optimal Solutions of the Hybrid LQG problem)}}
For the system governed by \eqref{LinearSDE}-\eqref{HybridCostLQG}, assume that a family of matrices $\left\{\Pi^{Q_j}\left(t\right); j = 0,1,\cdots,L \right\}$ exists 
%and $\Pi^{Q_j} \equiv \Pi^{Q_j}\left(t\right)$ 
satisfying the following family of Riccati equations (for simplicity of notation, the explicit time dependence  $\left(t\right)$ is dropped whenever it is clear from the context)
\begin{multline} 
\dot{\Pi}^{Q_{j}}=\Pi^{Q_{j}}B^{Q_{j}}\left[R^{Q_{j}}\right]^{-1} \left[B^{Q_{j}}\right]^T \Pi^{Q_{j}}-\Pi^{Q_{j}}A^{Q_{j}}\allowdisplaybreaks\\-[A^{Q_{j}}]^{T}\Pi^{Q_{j}}-P^{Q_{j}}, \label{StochRiccatiDynamics}
\end{multline}
%such that
subject to the terminal and boundary conditions
\begin{align}
\Pi^{Q_L}\left(T\right) &= \bar{P}^{Q_L}, \label{StochasticRiccatiTerminal} 
\allowdisplaybreaks\\
\Pi^{Q_{j-1}}\left(t_{j}\right) &= \Psi_{\sigma_{j}}^{T}\Pi^{Q_{j}}\left(t_{j}\right)\Psi_{\sigma_{j}} + C_{\sigma_j}(t_j), \label{StochasticRiccatiBC}
\end{align}
and for every $j=L, L-1,\cdots,1$, %(determined from a backward sequence), 
there exist $t_j \in \left[0,t_{j+1}\right)$ %${\cal F}^t$-independent state-invariant (almost surely deterministic) solution of 
satisfying the following algebraic matrix relations (equality, strict positive definiteness,  and strict negative definiteness):
\begin{align}
H^{\Delta}_{\sigma_j}\left(s\right) &= 0, \hspace{30pt} {s=t_{j}}, \label{DeltaHzero}
\\
H^{\Delta}_{\sigma_j}\left(s\right) &> 0 , \hspace{30pt} s>t_{j}, \label{DeltaHpositive}
\\
H^{\Delta}_{\sigma_j}\left(s\right) &< 0 , \hspace{30pt} s<t_{j}, \label{DeltaHnegative}
\end{align}
where the time order of the strict matrix inequalities corresponds to the strict decrease in the value function, and 
\begin{multline}
H^{\Delta}_{\sigma_j}\left(s\right) := \Psi_{\sigma_{j}}^{T}\Pi^{Q_{j}}(s)\bigg[B^{Q_{j}}[R^{Q_{j}}]^{-1} [B^{Q_{j}}]^{T}\allowdisplaybreaks\\-\Psi_{\sigma_{j}} B^{Q_{j-1}} [R^{Q_{j-1}}]^{-1} [B^{Q_{j-1}}]^{T}\Psi_{\sigma_{j}}^{T}\bigg]\Pi^{Q_{j}}(s)\Psi_{\sigma_{j}}
\allowdisplaybreaks\\
+\Psi_{\sigma_{j}}^{T}\Pi^{Q_{j}}(s)\bigg[\Psi_{\sigma_{j}}A^{Q_{j-1}}-A^{Q_{j}}\Psi_{\sigma_{j}}-\Psi_{\sigma_{j}}B^{Q_{j-1}}\allowdisplaybreaks\\[-1pt] [R^{Q_{j-1}}]^{-1} [B^{Q_{j-1}}]^{T} C_{\sigma_{j}}\bigg]
\allowdisplaybreaks
+\bigg[[A^{Q_{j-1}}]^{T} \Psi_{\sigma_{j}}-\Psi_{\sigma_{j}}^{T} [A^{Q_{j}}]^{T}\allowdisplaybreaks\\[-1pt]
-C_{\sigma_{j}} B^{Q_{j-1}} [R^{Q_{j-1}}]^{-1} [B^{Q_{j-1}}]^{T} \Psi_{\sigma_{j}}^{T}\bigg] \Pi^{Q_{j}}(s)\Psi_{\sigma_{j}}
\allowdisplaybreaks\\
+P^{Q_{j-1}}-C_{\sigma_{j}}B^{Q_{j-1}}[R^{Q_{j-1}}]^{-1} [B^{Q_{j-1}}]^{T} C_{\sigma_{j}} +C_{\sigma_{j}}A^{Q_{j-1}}\allowdisplaybreaks\\+[A^{Q_{j-1}}]^{T} C_{\sigma_{j}}-\Psi_{\sigma_{j}}^{T} P^{Q_{j}}\Psi_{\sigma_{j}} - \frac{\partial{C}_{\sigma_j}(t)}{\partial t}{\bigg|}_{t=s} . \label{StochHamiltonianJumpPathIndependent}
\end{multline} 
Then the switching times are ${\cal F}_t$-independent (almost surely deterministic) and are independent of the initial conditions, and the associated optimal control actions are determined as
\begin{equation}
u^{Q_j, \circ}\left(t,x\right)=-\left[R^{Q_j}\left(t\right)\right]^{-1} \left[B^{Q_j}\left(t\right)\right]^T \Pi^{Q_j} \left(t\right) x^{Q_j,\circ} \left(t\right).
\label{OptimalFeedbackLaw_switch}
\end{equation}
\hfill $\square$
\end{thm}
\textit{Proof}. See Appendix \ref{proofThm1}.

An important consequence of Theorem \ref{thm:HybridLQG} is that it yields as a corollary the crucial existence condition for the optimal stopping times used in Theorem \ref{thm:ENE_hybrid}.

Consider a system governed by
\begin{equation}
dx(t)=\left(A\left(t\right) x\left(t\right) +B\left(t\right)u\left(t\right)\right)dt +D\left(t\right)dw(t), 
\label{LinearSDEStopping}
\end{equation}
where $t\in\left[0,t^{\omega}_{s}\right)$, and $t^{\omega}_{s}$ is an ${\cal F}_t$-adapted stopping time, to be determined together with a continuous input in order to infimize (minimize) the cost
\begin{equation}
J(u)=  \frac{1}{2}  \mathbb{E}\Bigg[ {\Vert x(t_{s}^{\omega})\Vert^{2}_{C(t_s^\omega)}}
+ \int_0^{{t_{s}^{\omega}}}\left(\Vert x(t)\Vert_{P(t)}^{2}+\Vert u(t)\Vert_{R(t)}^{2}\right)dt \Bigg],
\label{StoppingCost}
\end{equation}
Define
\begin{multline}
H^{\Delta}\left(s\right):=P\left(s\right)-C\left(s\right)B\left(s\right)R^{-1}\left(s\right)B^{T}\left(s\right)C\left(s\right)\\+C\left(s\right)A\left(s\right)+ A^{T}\left(s\right)C\left(s\right)-\left.\frac{\partial C(t)}{\partial t}\right|_{t=s}. \label{DeltaHamiltonianStopping}
\end{multline}

\begin{cor}[{\small Stopping Policies for LQG Systems}]\label{thm:HybridLQGstop}
Consider the (deterministic) algebraic matrix expression \eqref{DeltaHamiltonianStopping}. If there exists a finite time $t_s\in \left[0,\infty\right)$ for which
\begin{align}  
H^{\Delta}\left(s\right) &= 0, \hspace{30pt} {s=t_{s}}, \label{stopHzero} 
\\
H^{\Delta}\left(s\right) &> 0 , \hspace{30pt} s>t_{s}, \label{stopHpositive}
\\
H^{\Delta}\left(s\right) &< 0 , \hspace{30pt} s<t_{s}, \label{stopHnegative}
\end{align}
then $t^{\omega}_{s} =t_{s}$ for almost all $\omega \in \Omega$, that is to say, the optimal stopping time for the system \eqref{LinearSDEStopping} with the cost \eqref{StoppingCost} is ${\cal F}_t$-independent, state-invariant, and takes the value $t_s$ almost surely, and the optimal input is determined by
\begin{equation}
u\left(t,x\right)=-R^{-1}\left(t\right)B^T\left(t\right) \Pi \left(t\right) x \left(t\right),
\label{OptimalFeedbackLaw}
\end{equation}
\\[-4pt]
\noindent where $\Pi \left(t\right)$ is the solution to
\begin{equation} 
\dot{\Pi}=\Pi  B  R^{-1}  B^T  \Pi -\Pi  A  -A^{T}  \Pi  - P  , \label{StochRiccatiDynamicsStopping}
\end{equation}
\\[-4pt]  
\noindent subject to the terminal (stopping) condition
\begin{equation} 
\Pi \left(t_{s}\right) = C \left(t_{s}\right). \label{StochasticRiccatiStopping}
\end{equation}
\hfill $\square$
\end{cor}
\textit{Proof.}
The proof is immediate since it expresses the conditions of Theorem \ref{thm:HybridLQG} for the special case of $\Psi_{\sigma} = 0$.
\section{Major-Minor Hybrid LQG Mean Field Games} %\vspace{-6pt}
\label{sec:ProblemFormulation}
\subsection{Problem Description} \label{sec:ProblemDescrb}
We consider the case where there exists one major agent and $N$ minor agents interacting with each other through the mean field coupling in their dynamics over the time interval $[0, T]$. Two types of minor agents are considered: type $\mathcal{A}^a$ with the population of $N_a$ and type $\mathcal{A}^b$ with the population of $N_b$, such that $N_a+N_b = N$.

The dynamics of the major agent and a generic minor agent are described by the linear time evolution of their states and a quadratic performance function. However, the two populations of minor agents have different linear dynamics and quadratic performance objectives. 
%In the scenario considered in this paper (see also Figure \ref{fig:HybridAutomata}) 
We study the case %interaction of agents over the interval $[0,T]$,
where the major agent $\mathcal{A}_0$ is permitted to switch from one set of dynamics to another at time $t_s^0$ if optimal, while a generic minor agent $\mathcal{A}_i,\, 1\leq i \leq N,$ is permitted to stop at an optimal time $ t_s^i$. With abuse of notation, the superscript $k$ in $\mathcal{A}_0^k,\, k=1,2$, denotes that the major agent's operation mode governed by the dynamics \eqref{MajorStandardDynamics_hybrid} and the cost functional \eqref{majorCostFunction_hybrid}, and in $\mathcal{A}_i^k,\,1 \leq i \leq N,\, k \triangleq a,b$, denotes that minor agent $\mathcal{A}_i,\, 1 \leq i \leq N$, is of type $k,\, k \triangleq a,b$,  governed by \eqref{MinorStandardDynamics_hybrid}-\eqref{minorCost_hybrid}. As discussed in Section \ref{sec:DiscreteState}, the optimal switching or stopping time policy for each agent is trajectory and state independent, and depends only on its dynamical parameters (i.e. the agent's type). Since the dynamical parameters for all minor agents in their respective types are the same, it follows that the stopping times are the same for all agents of each subpopulation. The distinct nature of the switching (stopping) events, together with the continuous evolution of the state processes between switchings, result in the stochastic hybrid form of the problem analyzed in this paper. Moreover, the fact that the minor agents are modeled as members of large populations gives rise to our use of the LQG MFG framework. The system has several distinct combinatoric alternatives; this is because there are various distinct sequences wherein one subpopulation of minor agents or another drops out first, or the major agent switches to one particular discrete state before or after a minor agent stopping event. It is to be emphasized that the discrete state sequence that actually occurs for any given system depends upon the solution of the complete (initial to terminal) hybrid MFG equations for the system, and in particular is not prescribed. We note that a key condition which yields the collective switching of the entire subpopulations is given by \eqref{A1EquivalentDiffusion} (see Section \ref{sec:HybridLQG}) and while this is reasonable in a class of LQG problems, the corresponding condition is most unlikely to hold in a nonlinear framework. 
%\vspace{-4pt}
\subsection{Discrete State Association}  %\vspace{-6pt}
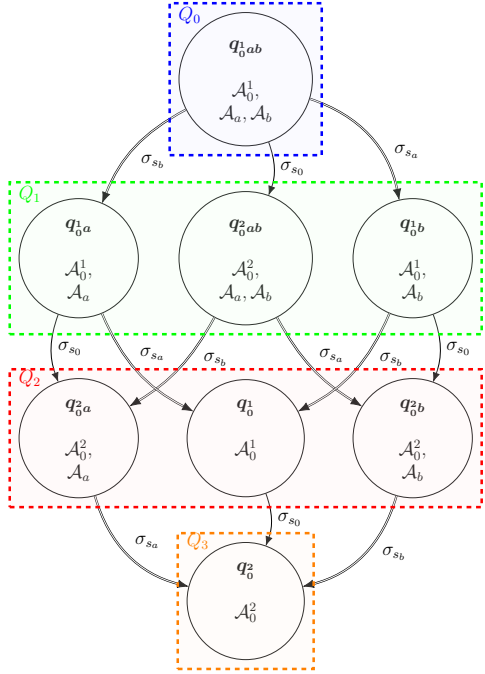
\begin{figure}
\centering
\scalebox{.35}{
\begin{tikzpicture}[node distance=37mm, auto]
  %%% first column %%%
  \node[state,minimum size=98pt](s123)
{$\begin{array}{c}
\mf{$\bm{q_{{}_0^1ab}}$} \\
\\
\mf{${\cal A}_{0}^{1}$},\\
\mf{${\cal A}_{a},{\cal A}_{b}$}
\end{array}$};
 
  %%% second column %%%
  \node[state,minimum size=98pt](s23) [right=15mm of s123]
{$\begin{array}{c}
\mf{$\bm{q_{{}_0^2ab}}$} \\
\\
\mf{${\cal A}_{0}^{2}$},\\
\mf{${\cal A}_{a},{\cal A}_{b}$}
\end{array}$};

  \node[state,minimum size=98pt](s13) [above of=s23]
{$\begin{array}{c}
\mf{$\bm{q_{{}_0^1a}}$} \\
\\
\mf{${\cal A}_{0}^{1}$},\\
\mf{${\cal A}_{a}$}
\end{array}$};

  \node[state,minimum size=98pt](s12) [below of=s23]
{$\begin{array}{c}
\mf{$\bm{q_{{}_0^1b}}$} \\
\\
\mf{${\cal A}_{0}^{1}$},\\
\mf{${\cal A}_{b}$}
\end{array}$};

  %%% third column %%%
  \node[state,minimum size=98pt] (s1) [right=17mm of s23]
{$\begin{array}{c}
\mf{$\bm{q_{{}_0^1}}$} \\
\\
\mf{${\cal A}_{0}^{1}$}\\{ }
\end{array}$};

\node[state,minimum size=98pt] (s2) [below of=s1]
{$\begin{array}{c}
\mf{$\bm{q_{{}_0^2b}}$} \\
\\
\mf{${\cal A}_{0}^{2}$},\\
\mf{${\cal A}_{b}$}
\end{array}$};

\node[state,minimum size=98pt] (s3) [above of=s1]
{$\begin{array}{c}
\mf{$\bm{q_{{}_0^2a}}$} \\
\\
\mf{${\cal A}_{0}^{2}$},\\
\mf{${\cal A}_{a}$}
\end{array}$};

  %%% fourth column %%%
  \node[state,minimum size=98pt](s0)[right=15mm of s1]
{$\begin{array}{c}
\mf{$\bm{q_{{}_0^2}}$} \\
\\
\mf{${\cal A}_{0}^{2}$}\\{ }
\end{array}$};

\path[-{Latex[scale=1.2]}]
(s123) edge[bend right, double, pos=0.75]  node {} (s12)
(s123) edge[bend left=20]  node {} (s23)
(s123) edge[bend left=20, double]  node {} (s13)

(s23) edge[bend right=15, double]  node {} (s2)
(s23) edge[bend left=15, pos=0.25, double]  node {} (s3)

(s12) edge[bend right=20]  node {} (s2)
(s12) edge[bend right=20, pos=0.25, double]  node {} (s1)

(s13) edge[bend left=20]  node {} (s3)
(s13) edge[bend left=20, pos=0.45, double]  node {} (s1)

(s1) edge[bend left=20]  node {} (s0)
(s2) edge[bend right, double, pos=0.35]  node {} (s0)
(s3) edge[bend left, double]  node {} (s0)

;

%\coordinate (q0) at ($(s123) + (0,0)$);
\node (A) [draw=blue, fit= (s123), inner sep=2mm, 
            dashed, ultra thick, fill=blue!10, fill opacity=0.2] {};
\node [xshift=3ex, yshift=10ex, blue] at (A.west) {\mf{$Q_0$}};

\node (B) [draw=myGrn, fit= (s23) (s12) (s13), inner sep=2mm, 
            dashed, ultra thick, fill=myGrn!10, fill opacity=0.2] {};
\node [xshift=3ex, yshift=33ex, myGrn] at (B.west) {\mf{$Q_1$}};
    
\node (C) [draw=red, fit= (s3) (s1) (s2), inner sep=2mm, 
            dashed, ultra thick, fill=red!10, fill opacity=0.2] {};
\node [xshift=3ex, yshift=33ex, red] at (C.west) {\mf{$Q_2$}};

\node (D) [draw=orange, fit= (s0), inner sep=2mm, 
            dashed, ultra thick, fill=orange!10, fill opacity=0.2] {};
\node [xshift=3ex, yshift=10ex, orange] at (D.west) {\mf{$Q_3$}};

\end{tikzpicture}
}
\caption{Hybrid Automata Diagram with a single major player and two populations of minor players with switching and stopping times. Transitions accompanied by dimension changes are identified with double-line arrows.} \label{fig:HybridAutomata}
%\vspace{-6pt}
\end{figure}

\label{sec:DiscreteState}
In order to present the dynamics of the system in the stochastic hybrid systems framework of \cite{APPECCDC2016, APPEC2017IFAC}, the discrete states $q_{{}_0^k\bullet}$ are assigned (see Figure \ref{fig:HybridAutomata}) where $k\triangleq a,b$ refers to the mode in the dynamics of the major agent and $\bullet$ represents the active subpopulations of minor agents. For instance, the discrete state $q_{{}_0^1ab}$ indicates that the major agent is subject to its first dynamics and both subpopulations $\mathcal{A}^a$ and $\mathcal{A}^b$ are present, and the discrete state $q_{{}_0^2a}$ indicates that the major agent is subject to its second dynamics, subpopulation $\mathcal{A}^a$ is present and subpopulation $\mathcal{A}^b$ has already quit the system. Furthermore, in order to refer to the temporal mode of the system, the multivalued discrete states $Q_j,~ 0\leq j \leq 3$, are introduced (see Figure \ref{fig:HybridAutomata}), which correspond to the evolution of the system within the intervals $[t_j,t_{j+1})$, where $t_0=0$ is the initial time, $t_1$, $t_2$, $t_3$ correspond to the times of the events of stopping of a subpopulation or switching of the major agent, in the order of occurrence, and $t_4=T$ is the terminal time. This corresponds to the scenario in which all the possible discrete changes in the system occur before the terminal time, i.e. $Q_3 = q_{{}_0^2}$. Other scenarios where the discrete state at terminal time is different from the case considered here are possible with minor variations over the results presented in this paper.

%\ali{I believe that with Section 2 preceding this discussion, the remark can be completely removed}
%{\color{blue}We remark that the hybrid MFG problems studied in this paper lie within the class of hybrid LQG problems for which optimal switching strategies are $\mathcal{F}_t$-independent, where $\mathcal{F}_t$ is the natural filtration associated with the sigma-algebra generated by the corresponding Wiener process (see Section \ref{sec:HybridLQG}). Therefore optimal switching or stopping strategies of an agent only depend on its dynamical parameters and hence coincide with those of all agents in its subpopulation.}

Now, we describe the evolution of the system over the sequence of generic discrete states $Q_j,\, 0 \leq j \leq 3$. The discrete state $Q_0$, as indicated in Figure \ref{fig:HybridAutomata}, associates with the system evolution over the interval $[0, t_1)$ in the system's initial setting where both subpopulations of minor agents are interacting together and with the major agent which is subject to its first dynamics $\mathcal{A}_0^1$. 

The multivalued discrete state $Q_1$ corresponds to the evolution of the system over $[t_1, t_2)$ with one change relative to the initial setting; this consists of three possible situations: (i) the major agent subject to its second dynamics $\mathcal{A}_0^2$ is interacting with both subpopulations $\mathcal{A}^a$, $\mathcal{A}^b$ present in the system; this corresponds to the centre node inside $Q_1$ in Figure \ref{fig:HybridAutomata} and is denoted by $Q_1 = q_{{}_0^2ab}$, (ii) the major agent subject to its first dynamics $\mathcal{A}_0^1$ is interacting with the subpopulation $\mathcal{A}^a$ while the subpopulation $\mathcal{A}^b$ has quit the system; this corresponds to the top node inside $Q_1$ in Figure \ref{fig:HybridAutomata} and is denoted by $Q_1 = q_{{}_0^1a}$, and (iii) the major agent subject to its first dynamics $\mathcal{A}_0^1$ is interacting with $\mathcal{A}^b$ while $\mathcal{A}^a$ has quit, corresponding to the bottom node inside $Q_1$ in Figure~\ref{fig:HybridAutomata}, denoted by $Q_1 = q_{{}_0^1b}$.

The multivalued discrete state $Q_2$ represents the evolution of the system over $[t_2, t_3)$ with two changes relative to the initial setting for which three situations can be considered: (I) the major agent subject to its second dynamics $\mathcal{A}_0^2$ is interacting with the subpopulation $\mathcal{A}^a$, and the subpopulation $\mathcal{A}^b$ has already quit, which corresponds to the top node inside $Q_2$ in Figure \ref{fig:HybridAutomata} denoted as $Q_2 = q_{{}_0^2a}$, (II) the major agent subject to its second dynamics $\mathcal{A}_0^2$ is interacting with $\mathcal{A}^b$, and the subpopulation $\mathcal{A}^a$ has already quit,  which corresponds to the bottom node inside $Q_2$ in Figure \ref{fig:HybridAutomata} denoted by $Q_2 = q_{{}_0^2b}$, (III) the major agent is subject to its first dynamics $\mathcal{A}_0^1$ and both subpopulations $\mathcal{A}^a$, $\mathcal{A}^b$ have already quit, which corresponds to the centre node \mbox{inside $Q_2$ in Figure \ref{fig:HybridAutomata}, denoted by $Q_2 = q_{{}_0^1}$}. 

The discrete state $Q_3$ corresponds to the evolution of the major agent subject to its second dynamics $\mathcal{A}_0^2$ over $[t_3, T]$ which corresponds to $Q_3 = q_{{}_0^2}$.

In this work it is assumed that each of the time periods $[ t_j, t_{j+1})$ associated with the multivalued discrete state $Q_j, \, 0\leq j \leq 3$, is non-empty, i.e. $t_{j} < t_{j+1}$. This assumption is tenable since it will be shown that for the class of hybrid LQG systems in this paper, the switching times $t_1,\,t_2,\, t_3$ can be deterministically evaluated as they depend only on the system parameters.      
\begin{remark} It would be possible to extend the formulation to include a larger number of subpopulations greater than two. However, this results in a larger number of discrete states and associated realizations for the system. Therefore the corresponding automata becomes more complex and the computational load increases. The problem may become intractable if the number of subpopulations becomes large.
\end{remark}
\subsection{Dynamics and Costs: Finite Population }
\label{sec:FinitePopulation}
%\vspace{-6pt}
\subsubsection{Major Agent}
%\vspace{-6pt}
Let the evolution of the major agent $\mathcal{A}_0^k,~ k= 1,2$, be expressed as
\begin{align}\label{MajorStandardDynamics_hybrid}
dx_0 = A_0^k x_0 dt + B_0^k u_0 dt + F_0^k x^{(N_t)} dt + D_0^k dw_0, 
\end{align}
where $x_0 \in \mathbb{R}^n$ is the state, $u_0 \in \mathbb{R}^m$ is the control input, and $w_0 \in \mathbb{R}^r$ is a standard Wiener process. The matrices $A_0^k$, $B_0^k$, $F_0^k$, and $D_0^k$, $ k=1,2$, are of appropriate dimension. 
 
From \eqref{MajorStandardDynamics_hybrid}, the major agent is coupled with the minor agents by the average term $x^{(N_t)} = \frac{1}{N_t} \sum_{i=1}^{N_t} x_i $. Note that in \eqref{MajorStandardDynamics_hybrid}, $N_t$ may take the following values. 
\begin{align} \label{NtCases_hybrid}
N_t = \begin{cases}
 N_a+N_b & \text{for } Q_0= q_{{}_0^1ab}, Q_1 = q_{{}_0^2ab} \\
 N_a &  \text{for } Q_1 = q_{{}_0^1a}, Q_2= q_{{}_0^2a}\\
 N_b & \text{for } Q_1 = q_{{}_0^1b}, Q_2 = q_{{}_0^2b} \\
 0 & \text{for } Q_2 =q_{{}_0^1}, Q_3 = q_{{}_0^2},
 \end{cases} 
 \end{align}
where $x^{(N_t)}:=0$ for $N_t=0$. The major agent $\mathcal{A}_0^k, k=1,2$, aims to minimize the following cost functional %\vspace{-5pt}
 \begin{gather} 
 J_0^{N,k}(u_0, u_{-0}) = \hf\mathbb{E} \big [ \Vert x_0(T) \Vert^2_{\bar{P}_0^k} \hfill  \nonumber\\ + \int_{0}^{T}  ( \Vert x_0 - \Phi(x^{(N_t)}) \Vert_{P_0^k}^2 + \Vert u_0 \Vert^2_{R_0^k} ) dt \big ], \label{majorCostFunction_hybrid}\\
 \Phi(x^{(N_t)}) \coloneqq H_0^k x^{(N_t)},
 \end{gather} %\vspace{-3pt}
with the weight matrices of appropriate dimension.
\begin{assum} (i) $(P^k_0)^\intercal = P^k_0$, $(R^k_0)^\intercal = R^k_0$, $(\bar{P}_0^k)^\intercal = \bar{P}_0^k$, \\(ii) the matrices $A_0^k$, $B_0^k$, $F_0^k$, $P^k_0$, $H^k_0$ and $R^k_0$ are bounded, (iii) $R^k_0$ is a continuous function of time $t$, $0\leq t \leq T$, and \\(iv) every column $D^k_0(:,j),\, j=1,\dots, r,$ of $D^k_0$ is such that 
$\int_0^T \Vert D^k_0(:,j) \Vert^2 < \infty$.\label{ass:major_parameters}
 \end{assum}
 \begin{assum}[Convexity] $R_0^k >0$, $\bar{P}_0^k \geq 0$, and $P_0^k \geq 0$.\label{ass:convexity_major}
 \end{assum} 
%Equation \eqref{MajorStandardDynamics_hybrid} together with the cost functional \eqref{majorCostFunction_hybrid} form a stochastic LQG problem for the major agent.  
We note once again that the superscript $k$ in $\mathcal{A}_0^k$ denotes that the major agent is acting with respect to the dynamics and the cost functional $k$. 
\subsubsection{Generic $\mathcal{A}^k$-Type Minor Agent} %\vspace{-6pt}
The dynamics for a minor agent $\mathcal{A}_i^k, k \triangleq a, b$, is given by 
 \begin{align}\label{MinorStandardDynamics_hybrid}
\hspace{-2.6mm} dx_i = A_k x_i dt + B_k u_i dt + G_k x_0 dt + F_k x^{(N_t)} dt + D_k dw_{i},
\end{align}
where $x_i \in \mathbb{R}^n$ is the state of agent $\mathcal{A}_i^k$, $u_i \in \mathbb{R}^m$ is the control input, $w_i \in \mathbb{R}^r$ is a standard Wiener process, and $A_k$, $B_k$, $G_k$, $F_k$, $D_k$ are constant matrices of appropriate dimension.    
\begingroup
Note that $N_t$ in \eqref{MinorStandardDynamics_hybrid}  again takes values as in \eqref{NtCases_hybrid} over the horizon $[0,T]$. The cost for a type $\mathcal{A}^k$ minor agent is given by %\vspace{-3pt}
\begin{gather} 
J_i^{N,k}(u_i, u_{-i}) = \hf\mathbb{E} \big [ \Vert x_i(t_s^i) - \Psi_k(x^{(N_{t_s^i})}) \Vert^2_{\bar{P}_k} \nonumber\\ + \int_{0}^{t_s^i}  ( \Vert x_i - \Psi_k(x^{(N_t)}, x_0) \Vert_{P_k}^2 + \Vert u_i \Vert^2_{R_k} ) dt \big ],  \label{minorCost_hybrid}\\
\Psi_k(x^{(N_t)}, x_0) \coloneqq H_1^k x_0(t) + H_2^k x^{(N_{t})},
\end{gather}
where the weight matrices have appropriate dimensions, and $t_s^i \in [0,T]$ is the stopping time of agent $i$ freely decided by this agent in order to minimize its individual cost. Particularly, $t_s^i$ is not directly restricted to be the same over the entire population. However, whenever the parameters of the associated extended dynamics (presented in Section \ref{sec:MinorInf}) satisfy the requirements of Corollary \ref{thm:HybridLQGstop}, then the optimal stopping times $t_s^i = t_s^k$ for the entire subpopulation of $\mathcal{A}^k$-type minor agents.% (similarly for $\mathcal{A}^b$-type minor agents in the next section). 
\begin{assum} (i) $P_k^\intercal = P_k$, $R_k^\intercal = R_k$, $\bar{P}_k^\intercal = \bar{P}_k$, (ii) the matrices $A_k$, $B_k$, $F_k$, $G_k$ ,$P_k$, $\bar{P}_k$, $H_1^k$, $H_2^k$ and $R_k$ are bounded, (iii) $R_k$ is a continuous function of time $t$, $0\leq t \leq T$, and (iv) every column $D_k(:,j),\, j=1,\dots, r,$ of $D_k$ is such that 
$\int_0^T \Vert D_k(:,j) \Vert^2 < \infty$.\label{ass:minor_parameters}
 \end{assum} 
\begin{assum}[Convexity] $\bar{P}_k \geq 0$, $P_k \geq 0 $, and $R_k > 0$.\label{ass:convexity_minor}
\end{assum} 
%The set of equations (\ref{MinorAcqStandardDynamics_hybrid}) and (\ref{minorAcqCost_hybrid}) constitute the stochastic optimal control problem for a minor agent of type $\mathcal{A}^k$. 
From \eqref{MinorStandardDynamics_hybrid} and \eqref{minorCost_hybrid}, a generic $\mathcal{A}^k$-type minor agent interacts with the major agent's state as well as the average state of all existing minor agents through its dynamics and cost functional.
\endgroup

We denote by $w=\{w_i, 0\leq i \leq N \}$ the set of $(N + 1)$ independent $\mb{R}^r$-valued standard Wiener processes on the probability space $(\Omega, \mc{F}, P)$, where $w$ is progressively measurable with respect to the filtration $\mc{F}^w=\{\mc{F}^w_t \subset \mc{F}; t \geq 0\}$.

\begin{assum} \label{IntialStateAss_HybridMFG}
The initial states $\lbrace x_i(0),~ 0 \leq i \leq N\rbrace$ defined on $(\Omega, \mathcal{F}, P)$ are identically distributed, mutually independent and also independent of $\mathcal{F}_{T}^{w}$, with $\mathbb{E}x_i(0)=0$. Moreover, $\sup_{i} \mathbb{E}\Vert x_i(0)\Vert^2 \leq c < \infty $, $0 \leq i \leq N<\infty$, with $c$ independent of $N$.
\end{assum} 
We also define
\begin{equation*}
\mathcal{I}_k := \{ 1 \leq i \leq N | \text{ agent $i$ belongs to sub-population $k$}\},
\end{equation*}
where $N_k := |\mathcal{I}_k\rvert$ is the number of agents in class $k\triangleq a, b$, with $ N_a + N_b = N$. The empirical distribution of the agents sampled independently of the initial conditions and Wiener processes within populations $\mathcal{A}^a$ and $\mathcal{A}^b$ at time $t_0$ is denoted by $\pi^{N} = (\pi_{a}^{N},\pi_{b}^N)$, where $ \pi_a^N = \tfrac{N_a}{N}$ and $ \pi_b^N = \tfrac{N_b}{N}$.

\begin{assum} \label{EmpiricalDistAss_HybridMFG}
There exists $\pi = (\pi_a , \pi_b)$ such that \linebreak $\mbox{lim}_{N \rightarrow \infty} \pi^N \stackrel{a.s.}{=} \pi $. 
\end{assum}
In the following we introduce the admissible sets of controls for each agent. {\color{black}By definition $\mc{F}_{i,t}, 1\leq i \leq N$, is the increasing family of null set augmented $\sigma$-fields generated by $(x_0(\tau), x_i(\tau); 0 \leq \tau \leq t)$}, and $\mc{F}_{0,t}$ is the increasing family of null set augmented $\sigma$-fields generated by $(x_0(\tau); 0 \leq \tau \leq t)$. $\mc{F}_t^N$ is the increasing family of $\sigma$-fields generated by the set $\{x_j(\tau), x_0(\tau); 0 \leq \tau \leq t,1 \leq j \leq N\}$. The set of control actions $\mathcal{U}^{N,L}_g$ consists of linear feedback control actions adapted to $\{\mc{F}_t^N , t \in [0,T]\}, 1 \leq N < \infty$.

\begin{assum}[\small{Major Agent Linear Controls}]\label{majorObs_HybridMFG}  For the major agent $\mathcal{A}_0$ the set of control inputs $\mathcal{U}_{0}^L$ is defined to be the collection of linear feedback controls adapted to the filtration $\lbrace \mathcal{F}_{0,t},\, t \in [0, T] \rbrace$. 
\end{assum}

\begin{assum}[\small{Minor Agent Linear  Controls}]\label{minorObs_HybridMFG}  For the minor agent $\mathcal{A}_i, 1 \leq i \leq N$, the set of control inputs $\mathcal{U}_{i}^L$ is defined to be the collection of linear feedback controls adapted to the filtration $\lbrace \mathcal{F}_{i,t},\, t \in [0, t_s^i] \rbrace, 1 \leq i \leq N$. 
\end{assum}
\section{Hybrid Mean Field Game Approach} \label{sec:InfinitePopulation}
%\vspace{-4pt}
Following the mean field game methodology with a major agent \cite{Huang2010, NourianSiam2013}, the hybrid MFG problem is first solved in the infinite population limit where the average term in the finite population dynamics and cost functional of each agent is replaced by its infinite population limit, i.e. the mean field,  and the major agent and a generic minor agent $\mathcal{A}_i$ only use local information (i.e. $\mc{F}_{0,t}$, $\mc{F}_{i,t}$, respectively.). Then specializing to linear systems (see e.g. \cite{Huang2010}), the major agent's state is extended with the mean field, while the minor agent's state is extended with the mean field and the major agent's state; this yields hybrid LQG optimal control problems (see appendix A) for each agent linked only through the mean field and the major agent's state. Then the main results of \cite{Huang2010}, \cite{NourianSiam2013} are (i) the existence of infinite population best response strategies which yield the Nash equilibria, and (ii) the infinite population best response strategies using local information applied to the finite population system yield an $\epsilon$-Nash equilibrium (see \textit{Theorem \ref{thm:ENE_hybrid}}). %{\color{blue}This indicates that if an agent unilaterally deviates from the obtained set of best-response strategies and employs an arbitrary strategy using information of all agents (i.e. adapted to $\mc{F}_t^N$), they can at most benefit by $\epsilon$, where $\epsilon$ is of order $\tfrac{1}{\sqrt{N}}$.} 

In this section, first, the hybrid evolution of the mean field is derived. Then the extended hybrid optimal control problems for the major agent and minor agents are formed and addressed in the infinite population case. Finally, \textit{Theorem \ref{thm:ENE_hybrid}} is presented which links the infinite population and finite population hybrid LQG MFG problem solutions.    

\subsection{Hybrid Evolution of Mean Field}
\label{sec:MF}
Following the LQG MFG methodology \cite{Huang2010}, the mean field is defined as the limit (in quadratic mean), when it exists, of the average of minor agents' states when the population size goes to infinity
\begin{align*}
\bar{x}^k(t) = \lim_{N_k \rightarrow \infty} x^{(N_k)}(t) = \lim_{N_k\rightarrow\infty} \frac{1}{N_k} \sum_{i=1}^{N_k} x_i(t), \quad q.m.
\end{align*}
where $k \triangleq a,b,$ for the case considered in this paper. Now, the control strategy for each minor agent is considered to have the general linear state feedback form %\vspace{-3pt}
\begin{equation} \label{generalFeedback_hybrid}
u_i = L_1^k x_i + L_2^k x_0 + \sum_{j \neq i, j=1}^{N_t} \hspace{-2mm}L_3^{k} x_j + m_k, \quad i \in \mathcal{I}_k,
\end{equation}
for bounded time-varying matrices $L_1^k, \,L_2^k, \,L_3^k$, and $m_k$ of appropriate dimension (for notation brevity here and in the rest of the paper time arguments are dropped unless for clarity). Then the mean field dynamics, in q.m.,  is obtained by substituting (\ref{generalFeedback_hybrid}) in the minor agents' dynamics (\ref{MinorStandardDynamics_hybrid}), and taking the average over subpopulation $\mathcal{A}^k$, and then its limit as $N_{k}\rightarrow \infty$. \\
With the assignment of discrete states $Q_j$ introduced in Section \ref{sec:DiscreteState}, the set of the mean field equations 
is given by 
\begin{equation} \label{MFequation_hybrid}
d\bar{x}^{Q_j} = \bar{A}^{Q_j} \bar{x}^{Q_j}dt + \bar{G}^{Q_j} x_0^{Q_j} dt + \bar{m}^{Q_j} dt,\quad  j=0, 1, 2, 3. \hspace{-1pt}  \end{equation}
For $Q_0= q_{{}_0^1ab}$, $\bar{x}^{Q_0}=[\bar{x}_a^T, \bar{x}_b^T]^T$ consists of the mean field $\bar{x}_a$ of the subpopulation $\mathcal{A}^a$, and the mean field $\bar{x}_b$ of the subpopulation $\mathcal{A}^b$ with $\pi^{Q_0} = \pi.$
The matrices in \eqref{MFequation_hybrid} are then
\begin{align}\label{MeanFieldMatrices}
\bar{A}^{Q_0} = \begin{bmatrix}
\bar{A}_a \\ 
\bar{A}_b
\end{bmatrix}, \quad
 \bar{G}^{Q_0} = \begin{bmatrix}
\bar{G}_a \\
\bar{G}_b
\end{bmatrix}, \quad
 \bar{m}^{Q_0}= \begin{bmatrix}
 \bar{m}_a\\
 \bar{m}_b
\end{bmatrix},  
\end{align}
where $\bar{A}_a, \bar{A}_b \in \mathbb{R}^{n \times 2n}$, $\bar{G}_a, \bar{G}_b \in \mathbb{R}^{n \times n }$, $\bar{m}_a, \bar{m}_b \in \mb{R}^n$. The above matrices shall be determined from the consistency equations discussed in Section \ref{sec:ConsistencyCondition}. 

In case (i) in Section \ref{sec:DiscreteState} where $Q_1 = q_{{}_0^2ab}$, the mean field is defined as $\bar{x}^{q_{{}_0^2ab}}=[\bar{x}_a^T, \bar{x}_b^T]^T$, hence
$\pi^{q_{{}_0^2ab}} = \pi$, and
\begin{align}\label{MeanFieldMatrices}
\bar{A}^{q_{{}_0^2ab}} = \begin{bmatrix}
\bar{A}_a \\ 
\bar{A}_b
\end{bmatrix}, \quad
 \bar{G}^{q_{{}_0^2ab}} = \begin{bmatrix}
\bar{G}_a \\
\bar{G}_b
\end{bmatrix}, \quad
 \bar{m}^{q_{{}_0^2ab}}= \begin{bmatrix}
 \bar{m}_a\\
 \bar{m}_b
\end{bmatrix}.  
\end{align}
For case (ii) where $Q_1 = q_{{}_0^1a}$, $\bar{x}^{q_{{}_0^1a}}=\bar{x}_a$, and hence
$\pi^{q_{{}_0^1a}}= 1,$
and the matrices in \eqref{MFequation_hybrid} are given as 
\begin{align}
\bar{A}^{q_{{}_0^1a}} = \bar{A}_a, \quad \bar{G}^{q_{{}_0^1a}} = \bar{G}_a, \quad \bar{m}^{q_{{}_0^1a}}=\bar{m}_a,
\end{align}  
where $\bar{A}_a \in \mathbb{R}^{n \times n}$, $\bar{G}_a \in \mathbb{R}^{n \times n}$, $\bar{m}_a \in \mathbb{R}^{n}$.

For case (iii) where $Q_1 =q_{{}_0^1b}$, 
$\bar{x}^{q_{{}_0^1b}}=\bar{x}_b$, and hence 
$\pi^{q_{{}_0^1b}} = 1,$
and the matrices in \eqref{MFequation_hybrid} are given by
\begin{align}
\bar{A}^{q_{{}_0^1b}} = \bar{A}_b, \quad \bar{G}^{q_{{}_0^1b}} = \bar{G}_b, \quad \bar{m}^{q_{{}_0^1b}}=\bar{m}_b. 
\end{align} 
For case (I) in Section \ref{sec:DiscreteState} where $Q_2 = q_{{}_0^2a}$, the mean field is defined as $\bar{x}^{q_{{}_0^2a}} = \bar{x}_a$, and hence
$\pi^{q_{{}_0^2a}} = 1,$
and the matrices in \eqref{MFequation_hybrid} are given as
\begin{align}
\bar{A}^{q_{{}_0^2a}} = \bar{A}_a, \quad \bar{G}^{q_{{}_0^2a}} = \bar{G}_a, \quad \bar{m}^{q_{{}_0^2a}}=\bar{m}_a. 
\end{align}
For case (II) where $Q_2 = q_{{}_0^2b}$, $\bar{x}^{q_{{}_0^2b}} = \bar{x}_b$, and hence
$\pi^{q_{{}_0^2b}}= 1,$
and the matrices in \eqref{MFequation_hybrid} are given by
\begin{align}
\bar{A}^{q_{{}_0^2b}} = \bar{A}_b, \quad \bar{G}^{q_{{}_0^2b}} = \bar{G}_b, \quad \bar{m}^{q_{{}_0^2b}}=\bar{m}_b. 
\end{align}
For case (III) where $Q_2 = q_{{}_0^1}$, \mbox{$\bar{x}^{q_{{}_0^1}} = 0$, hence $\pi^{q_{{}_0^1}} =0 $}.

Finally, for $Q_3=q_{{}_0^2}$, $\bar{x}^{Q_3}=0$, and as a result $\pi^{Q_3}= 0 $.

\subsection{Major Agent: Infinite Populations} \label{sec:MajorInf}
%\vspace{-6pt}
\subsubsection{Hybrid Dynamics and Cost}\label{sec:hybDynCost}
The extended hybrid dynamics of the major agent  in the infinite population, i.e. the dynamics for $x_0^{ex, Q_j}$  is given by
\begin{equation}
dx_0^{ex, Q_j} = (\mathbb{A}_0^{Q_j} x_0^{ex,Q_j} + \mathbb{M}_0^{Q_j} + \mathbb{B}_0^{Q_j} u_0^{Q_j}) dt + \mathbb{D}_0^{Q_j} dW_0^{Q_j}, 
\label{MajorHybridDSE_hybrid}
\end{equation}
$0\leq j \leq 3$, where the dynamical matrices are given by
\begin{gather}
\mathbb{A}_0^{Q_j} = \begin{bmatrix}
A_0^{Q_j} & \pi^{Q_j} \otimes F_0^{Q_j}\\
\bar{G}^{Q_j} & \bar{A}^{Q_j}
\end{bmatrix}, ~
\mathbb{M}_0^{Q_j} = \begin{bmatrix}
0_{n \times 1} \\
\bar{m}^{Q_j}
\end{bmatrix}, ~
\mathbb{B}_0^{Q_j} = \begin{bmatrix}
B_0^{Q_j}\\
0_{\bullet \times \bullet}
\end{bmatrix}, \nonumber \allowdisplaybreaks\\
\mathbb{D}_0^{Q_j} = \begin{bmatrix}
D_0^{Q_j} & 0_{\bullet \times \bullet}\\
0_{\bullet \times \bullet} & 0_{\bullet \times \bullet} \end{bmatrix}, \quad 
W_0^{Q_j} = \begin{bmatrix}
w_0\\
0_{\bullet \times \bullet}
\end{bmatrix}. \label{majorExMatGenDyn}
\end{gather}
In \eqref{majorExMatGenDyn}, $0_{\bullet \times \bullet}$ denotes a zero matrix of appropriate dimension, and $\otimes$ denotes the Kronecker product. 

The cost functional for the extended major agent's hybrid system is given by
\begin{multline}
J_0^{\infty}(u_0)= \hf\mathbb{E} \Big [ \Vert x^{ex,Q_3}_0(T) \Vert^2_{\bar{\mathbb{P}}_0^{Q_3}} + \sum_{j=1}^{3} \Vert x_0^{ex,Q_j}(t_j^-)\Vert^2_{\mathbb{C}_{0,\sigma_j}} \allowdisplaybreaks\\ + \sum_{j=0}^{3} \int_{t_j}^{t_{j+1}} \big ( \Vert x^{ex,Q_j}_0(s) \Vert^2_{\mathbb{P}_0^{Q_j}} + \Vert u_0^{Q_j}(s) \Vert^2_{R_0^{Q_j}} \big ) ds \Big], \label{majorCostQuadratic_hybrid}
\end{multline}
where $t_0 = 0,\, t_4 = T$. In \eqref{majorCostQuadratic_hybrid}, the first term denotes terminal cost and the third term denotes running cost where the corresponding weight matrices are defined as 
\begin{gather}
\bar{\mathbb{P}}_0^{Q_3} = \bar{P}_0^2, \nonumber \allowdisplaybreaks\\
\mathbb{P}_0^{Q_j} = [I_n, -\pi^{Q_j}\otimes H_0^{Q_j}]^T P_0^{Q_j} [I_n, -\pi^{Q_j}\otimes H_0^{Q_j}], \label{majorExMatGenCost}
\end{gather}
where $I_n$ denotes the identity matrix of size $n$. Moreover, the second term in \eqref{majorCostQuadratic_hybrid} denotes the switching cost corresponding to the terminal cost of the quitting agents, %\ali{denotes switching cost "for the mean field that corresponds to the terminal cost of the quitting agents"} \dena{incorporated.}
for which the associated weight matrix $\mathbb{C}_{0,\sigma_j}$ shall be identified for each switching in Section \ref{sec:majorJumpMapNswitchingCost}. 

Now the dynamical and weight matrices introduced in their general form, respectively, in \eqref{majorExMatGenDyn} and \eqref{majorExMatGenCost} are specified for each discrete state $Q_j, \, 0 \leq j \leq 3$. 

Over the interval $\left[t_0,t_1\right)$, and in discrete state $Q_0$, the dynamics of the continuous state $x_0^{ex,Q_0}=[x_0^T, \bar{x}_a^T, \bar{x}_b^T]^T$ is determined by \eqref{MajorHybridDSE_hybrid} with
\begin{gather}
\mathbb{A}_0^{Q_0}\! =\! \begin{bmatrix}
A_0^1 &   \pi \otimes F_0^1  \\
\begin{bmatrix}
\bar{G}_a \\
\bar{G}_b \end{bmatrix}  & 
\begin{bmatrix}
\bar{A}_a\\
\bar{A}_b \end{bmatrix}
\end{bmatrix}, ~~
\mathbb{M}_0^{Q_0} \!=\! \begin{bmatrix} 
0_{n \times 1} \\
\begin{bmatrix}
\bar{m}_a\\
\bar{m}_b \end{bmatrix}
\end{bmatrix}, ~ 
W_0^{Q_0}\! =\! \begin{bmatrix}
w_0\\
0_{2r \times 1}
\end{bmatrix}, \nonumber\allowdisplaybreaks\\ \mathbb{B}_0^{Q_0} \!=\! \begin{bmatrix}
B_0^1 \\
0_{2n \times m}
\end{bmatrix}, ~~
\mathbb{D}_0^{Q_0} \!=\! 
 \begin{bmatrix}
D_0^1 & 0_{n \times 2r}\\
0_{2n \times r} & 0_{2n \times 2r}
\end{bmatrix}, 
\end{gather}
where $\pi \otimes  F_0^1 =  [\pi_aF_0^1, \pi_bF_0^1]$, and $\mathbb{P}_0^{Q_0}$
in \eqref{majorCostQuadratic_hybrid} is given by
\begin{align}
\mathbb{P}_0^{Q_0} = [I_n,-\pi_a H_0^{1},-\pi_b H_0^{1}]^T P_0^{1} [I_n, -\pi_a H_0^{1},-\pi_b H_0^{1}].
\end{align}
We also define 
\begin{equation}
\bar{\mathbb{P}}_0^{Q_0} = [I_n, -\pi_a H_0^{1},-\pi_b H_0^{1}]^T \bar{P}_0^{1} [I_n, -\pi_a H_0^{1},-\pi_b H_0^{1}], \label{Q0SCost}
%\nonumber \\&= 
%\left[\begin{array}{ccc}
%\bar{\mathbb{P}}_{0,11}^{Q_0} & \bar{\mathbb{P}}_{0,12}^{Q_0} & \bar{\mathbb{P}}_{0,13}^{Q_0}\\
%\bar{\mathbb{P}}_{0,21}^{Q_0} &
%\bar{\mathbb{P}}_{0,22}^{Q_0} & 
%\bar{\mathbb{P}}_{0,23}^{Q_0}\\
%\bar{\mathbb{P}}_{0,31}^{Q_0} &
%\bar{\mathbb{P}}_{0,32}^{Q_0} &
%\bar{\mathbb{P}}_{0,33}^{Q_0}
%\end{array}\right],\quad \{\bar{\mathbb{P}}_{0,ij}^{Q_0},\, i,j=1,2,3 \} \in \mathbb{R}^n, 
\end{equation}
which will be used in section \ref{sec:majorJumpMapNswitchingCost} to specify the switching cost at $t_1$.  

Over the interval $\left[t_1,t_2\right)$, in case (i) where $Q_1 = q_{{}_0^2ab}$ holds over the interval $\left[t_1,t_2\right)$, the dynamics of $x_0^{ex,q_{{}_0^2ab}}=[x_0^T, \bar{x}_a^T, \bar{x}_b^T]^T$ is governed by \eqref{MajorHybridDSE_hybrid} with %\vspace{-6pt}
\begin{gather}
\mathbb{A}_0^{q_{{}_0^2ab}} = \begin{bmatrix}
A_0^2 & \pi \otimes F_0^2  \\
\begin{bmatrix}
\bar{G}_a \\
\bar{G}_b \end{bmatrix}  & 
\begin{bmatrix}
\bar{A}_a\\
\bar{A}_b \end{bmatrix}
\end{bmatrix},~ \mathbb{M}_0^{q_{{}_0^2ab}} = \begin{bmatrix} 
0_{n \times 1} \\
\begin{bmatrix}
\bar{m}_a\\
\bar{m}_b \end{bmatrix}
\end{bmatrix}, \nonumber \allowdisplaybreaks\\
\mathbb{B}_0^{q_{{}_0^2ab}} = \begin{bmatrix}
B_0^2 \\
0_{2n \times m}
\end{bmatrix}, ~
\mathbb{D}_0^{q_{{}_0^2ab}} = 
 \begin{bmatrix}
D_0^2 & 0_{n \times 2r}\\
0_{2n \times r} & 0_{2n \times 2r}
\end{bmatrix},~ W_0^{q_{{}_0^2ab}} = \begin{bmatrix}
w_0\\
0_{2r \times 1}
\end{bmatrix},
\end{gather}
 and $\mathbb{P}_0^{q_{{}_0^2ab}}$
in \eqref{majorCostQuadratic_hybrid} and $\bar{\mathbb{P}}_0^{q_{{}^2_0ab}}$ (to be used in Section \ref{sec:majorJumpMapNswitchingCost} for specifying the switching cost at $t_2$.) are given by
\begin{gather}
\mathbb{P}_0^{q_{{}_0^2ab}} =[I_n,-\pi_a H_0^{2},-\pi_b H_0^{2}]^{T}P_0^{2}[I_n, -\pi_a H_0^{2},-\pi_b H_0^{2}],\\
\bar{\mathbb{P}}_0^{q_{{}^2_0ab}} = [I_n, -\pi_a H_0^{2},-\pi_b H_0^{2}]^T \bar{P}_0^{2} [I_n, -\pi_a H_0^{2},-\pi_b H_0^{2}].  \label{q02abSCost}
\end{gather}
%Moreover, 
%\begin{align}
%\bar{\mathbb{P}}_0^{q_{{}^2_0ab}} &= [I_n, -\pi_a H_0^{2},-\pi_b H_0^{2}]^T %\bar{P}_0^{2} [I_n, -\pi_a H_0^{2},-\pi_b H_0^{2}],  \label{q02abSCost}
% \nonumber \\&= 
%\left[\begin{array}{ccc}
%\bar{\mathbb{P}}_{0,11}^{q_{{}^2_0ab}} & \bar{\mathbb{P}}_{0,12}^{q_{{}^2_0ab}} & \bar{\mathbb{P}}_{0,13}^{q_{{}^2_0ab}}\\
%\bar{\mathbb{P}}_{0,21}^{q_{{}^2_0ab}} &
%\bar{\mathbb{P}}_{0,22}^{q_{{}^2_0ab}} & 
%\bar{\mathbb{P}}_{0,23}^{q_{{}^2_0ab}}\\
%\bar{\mathbb{P}}_{0,31}^{q_{{}^2_0ab}} &
%\bar{\mathbb{P}}_{0,32}^{q_{{}^2_0ab}} &
%\bar{\mathbb{P}}_{0,33}^{q_{{}^2_0ab}}
%\end{array}\right],\quad \{\bar{\mathbb{P}}_{0,ij}^{q_{{}^2_0ab}},\, i,j=1,2,3 \} \in \mathbb{R}^n,
%\end{align}
%which will be used in Section \ref{sec:majorJumpMapNswitchingCost} to specify the switching cost at $t_2$. 
Over the interval $\left[t_1,t_2\right)$, in case (ii) where $Q_1 = q_{{}_0^1a}$ holds, the dynamics and cost functional for $x_0^{ex,q_{{}_0^1a}}=[x_0^T, \bar{x}_a^T]^T$ are, respectively, determined by \eqref{MajorHybridDSE_hybrid} and \eqref{majorCostQuadratic_hybrid} with
\begin{gather}
\mathbb{A}_0^{q_{{}_0^1a}} = \begin{bmatrix}
A_0^1 & F_0^1 \\
\bar{G}_a & \bar{A}_a
\end{bmatrix}, \quad \mathbb{M}_0^{q_{{}_0^1a}} = \begin{bmatrix} 
0_{n \times 1} \\
\bar{m}_a
\end{bmatrix}, \quad \mathbb{B}_0^{q_{{}_0^1a}} = \begin{bmatrix}
B_0^1 \\
0_{n \times m}
\end{bmatrix}, \nonumber \allowdisplaybreaks\\
\mathbb{D}_0^{q_{{}_0^1a}} = 
 \begin{bmatrix}
D_0^1 & 0_{n \times r}\\
0_{n \times r} & 0_{n \times r}
\end{bmatrix}, \quad
W_0^{q_{{}_0^1a}} = \begin{bmatrix}
w_0\\
0_{r \times 1} \end{bmatrix},\allowdisplaybreaks\\
\mathbb{P}_0^{q_{{}_0^1a}} =[I_n, -H_0^{1}]^{T}P_0^1[I_n, -H_0^{1}],\allowdisplaybreaks\\
\bar{\mathbb{P}}_0^{q_{{}_0^1a}} =[I_n, -H_0^{1}]^{T}\bar{P}_0^1[I_n, -H_0^{1}].  \label{q01aSCost}
\end{gather}
%and the cost functional is determined by \eqref{majorCostQuadratic_hybrid} with $\mathbb{P}_0^{q_{{}_0^1a}} =[I_n, -H_0^{1}]^{T}P_0^1[I_n, -H_0^{1}]$. In addition, matrix $\bar{\mathbb{P}}_0^{q_{{}_0^1a}}$  which shall be used in Section \ref{sec:majorJumpMapNswitchingCost} to identify the switching cost at $t_2$ is defined as 
%\begin{align}
%\bar{\mathbb{P}}_0^{q_{{}_0^1a}} &=[I_n, -H_0^{1}]^{T}\bar{P}_0^1[I_n, %-H_0^{1}].  \label{q01aSCost}
%\left[\begin{array}{cc}
%\bar{\mathbb{P}}_{0,11}^{q_{{}^1_0a}} & \bar{\mathbb{P}}_{0,12}^{q_{{}^1_0a}} \\
%\bar{\mathbb{P}}_{0,21}^{q_{{}^1_0a}} & \bar{\mathbb{P}}_{0,22}^{q_{{}^1_0a}} 
%\end{array}\right],\quad \{\bar{\mathbb{P}}_{0,ij}^{q_{{}^1_0a}},\, i,j=1,2 \} \in \mathbb{R}^n.
%\end{align}
Over the interval $\left[t_1,t_2\right)$, in case (iii) where $Q_1 = q_{{}_0^1b}$ holds, $x^{ex,q_{{}_0^1b}}=[x_0^T, \bar{x}_b^T]^T$ and
\begin{gather}
\mathbb{A}_0^{q_{{}_0^1b}} = \begin{bmatrix}
A_0^1 & F_0^1 \\
\bar{G}_b & \bar{A}_b
\end{bmatrix}, \quad \mathbb{M}_0^{q_{{}_0^1b}} = \begin{bmatrix} 
0_{n \times m} \\
\bar{m}_b
\end{bmatrix}, 
\quad
\mathbb{B}_0^{q_{{}_0^1b}} = \begin{bmatrix}
B_0^1 \\
0_{n \times m}
\end{bmatrix}, \nonumber \allowdisplaybreaks\\
\mathbb{D}_0^{q_{{}_0^1b}} = 
 \begin{bmatrix}
D_0^1 & 0_{n \times r}\\
0_{n \times r} & 0_{n \times r}
\end{bmatrix}, \quad
W_0^{q_{{}_0^1b}}= 
\begin{bmatrix}
w_0\\
0_{r \times 1}
\end{bmatrix}, \\
\mathbb{P}_0^{q_{{}_0^1b}} =[I_n, -H_0^1]^{T}P_0^1[I_n, -H_0^1],\allowdisplaybreaks\\
\bar{\mathbb{P}}_0^{q_{{}_0^1b}} =[I_n, -H_0^{1}]^{T}\bar{P}_0^1[I_n, -H_0^{1}].  \label{q01bSCost} 
%=\left[\begin{array}{cc}
%\bar{\mathbb{P}}_{0,11}^{q_{{}^1_0b}} & \bar{\mathbb{P}}_{0,12}^{q_{{}^1_0b}} \\
%\bar{\mathbb{P}}_{0,21}^{q_{{}^1_0b}} & \bar{\mathbb{P}}_{0,22}^{q_{{}^1_0b}} 
%\end{array}\right],
\end{gather}
%where $\{\bar{\mathbb{P}}_{0,ij}^{q_{{}^1_0a}},\, i,j=1,2 \} \in \mathbb{R}^n$.
Over the interval $\left[t_2,t_3\right)$, in case (I) where $Q_2 = q_{{}_0^2a}$ holds, $x^{ex,q_{{}_0^2 a}}=[x_0^T, \bar{x}_a^T]^T$ and
 \begin{gather}
\mathbb{A}_0^{q_{{}_0^2a}} = \begin{bmatrix}
A_0^2 & F_0^2 \\
\bar{G}_a & \bar{A}_a
\end{bmatrix}, \quad \mathbb{M}_0^{q_{{}_0^2a}} = \begin{bmatrix}
0_{n \times 1} \\
\bar{m}_a
\end{bmatrix}, \quad
\mathbb{B}_0^{q_{{}_0^2a}} = \begin{bmatrix}
B_0^2 \\
0_{n \times m}
\end{bmatrix}, \nonumber \allowdisplaybreaks\\
\mathbb{D}_0^{q_{{}_0^2a}} = 
 \begin{bmatrix}
D_0^2 & 0_{n \times r}\\
0_{n \times r} & 0_{n \times r}
\end{bmatrix}, \quad
W_0^{q_{{}_0^2a}} = \begin{bmatrix}
w_0\\
0_{r \times 1} \end{bmatrix}, \allowdisplaybreaks\\
\mathbb{P}_0^{q_{{}_0^2a}} =[I_n, -H_0^2]^{T}P_0^2[I_n, -H_0^2],\allowdisplaybreaks\\
\bar{\mathbb{P}}_0^{q_{{}_0^2a}} =[I_n, -H_0^{2}]^{T}\bar{P}_0^2[I_n, -H_0^{2}]. \label{q02aSCost}
%=\left[\begin{array}{cc}
%\bar{\mathbb{P}}_{0,11}^{q_{{}^2_0a}} & \bar{\mathbb{P}}_{0,12}^{q_{{}^2_0a}} \\
%\bar{\mathbb{P}}_{0,21}^{q_{{}^2_0a}} & \bar{\mathbb{P}}_{0,22}^{q_{{}^2_0a}} %\end{array}\right],
\end{gather}
%with $\{\bar{\mathbb{P}}_{0,ij}^{q_{{}^2_0a}},\, i,j=1,2 \} \in \mathbb{R}^n$.
Over the interval $\left[t_2,t_3\right)$, in case (II) where $Q_2 = q_{{}_0^2b}$ holds, $x^{ex,q_{{}_0^2b}}=[x_0^T, \bar{x}_b^T]^T$ and
 \begin{gather}
\mathbb{A}_0^{q_{{}_0^2b}} = \begin{bmatrix}
A_0^2 & F_0^2 \\
\bar{G}_b & \bar{A}_b
\end{bmatrix}, \quad \mathbb{M}_0^{q_{{}_0^2b}} = \begin{bmatrix} 
0_{n \times 1} \\
\bar{m}_b
\end{bmatrix}, \quad
\mathbb{B}_0^{q_{{}_0^2b}} = \begin{bmatrix}
B_0^2 \\
0_{n \times m}
\end{bmatrix}, \nonumber \allowdisplaybreaks\\
\mathbb{D}_0^{q_{{}_0^2b}} = 
 \begin{bmatrix}
D_0^2 & 0_{n \times r}\\
0_{n \times r} & 0_{n \times r}
\end{bmatrix}, \quad
W_0^{q_{{}_0^2b}} = \begin{bmatrix}
w_0\\
0_{r \times 1} \end{bmatrix}, \allowdisplaybreaks\\
\mathbb{P}_0^{q_{{}_0^2b}} =[I_n, -H_0^2]^{T}P_0^2[I_n, -H_0^2],\allowdisplaybreaks\\
\bar{\mathbb{P}}_0^{q_{{}_0^2b}} =[I_n, -H_0^{2}]^{T}\bar{P}_0^2[I_n, -H_0^{2}]. \label{q02bSCost}
%=\left[\begin{array}{cc}
%\bar{\mathbb{P}}_{0,11}^{q_{{}^2_0b}} & \bar{\mathbb{P}}_{0,12}^{q_{{}^2_0b}} \\
%\bar{\mathbb{P}}_{0,21}^{q_{{}^2_0b}} & \bar{\mathbb{P}}_{0,22}^{q_{{}^2_0b}} %\end{array}\right],
\end{gather}
%where $\{\bar{\mathbb{P}}_{0,ij}^{q_{{}^1_0a}},\, i,j=1,2 \} \in \mathbb{R}^n$.
Over the interval $\left[t_2 , t_3\right)$, in case (III) where $Q_2 = q_{{}_0^1}$ holds, $x^{ex,q_{{}_0^1}} = x_0$ and
\begin{gather*}
\mathbb{A}_0^{q_{{}_0^1}} = A_0^1, \quad 
\mathbb{M}_0^{q_{{}_0^1}} = 0_{n \times 1}, \quad
\mathbb{B}_0^{q_{{}_0^1}} = B_0^1, \quad 
\mathbb{D}_0^{q_{{}_0^1}} = D_0^1, \nonumber \allowdisplaybreaks\\
W_0^{q_{{}_0^1}} = w_0, \quad
\mathbb{P}_0^{q_{{}_0^1}} = P_0^1, \quad 
\bar{\mathbb{P}}_0^{q_{{}_0^1}} = \bar{P}^1_0. 
\end{gather*}
Finally, over the interval $\left[t_3,T\right]$, in discrete state $Q_3$, $x^{ex,Q_3} = x_0$  and 
\begin{gather*}
\mathbb{A}_0^{Q_3} = A_0^2, \quad 
\mathbb{M}_0^{Q_3} = 0_{n \times 1}, \quad
\mathbb{B}_0^{Q_3} = B_0^2, \quad 
\mathbb{D}_0^{Q_3} = D_0^2, \nonumber \allowdisplaybreaks\\
W_0^{Q_3} = w_0, \quad
\mathbb{P}_0^{Q_3} = P_0^2, \quad
\bar{\mathbb{P}}_0^{Q_3} = \bar{P}^2_0.
\end{gather*}
\subsubsection{Jump Transition Maps and Switching Costs} \label{sec:majorJumpMapNswitchingCost}
The major agent's switching cost associated with $t_j$ takes into account the cost incurred when a change occurs in the system. To identify it, we define the notation $M_0^{Q_j}(l:m), \, 0\leq j \leq 3$, which is formed by using matrix  $\bar{\mathbb{P}}^{Q_j}_0$ wherein all the entires are made zero except those associated with its $l$-th to $m$-th columns and rows. Hence it has the same dimension (size) as $\bar{\mathbb{P}}^{Q_j}_0$, i.e.    
%\vspace{0.15cm}
\begin{align}
%\overbrace{}^{\text{$l$ to $m$ th columns of} \, \, \bar{\mathbb{P}}_k^{Q_j} }
M_{0}^{Q_j}(l:m) = \left[ 
\begin{array}{c|c|c}
\bovermat{$\bar{\mathbb{P}}_0^{Q_j}(:,l:m)$}{0&\hspace{0.6cm}}&0 \\
\hline
%\vspace{0.01cm}
& &\\
\hline 
0&  &0
\end{array} \right]_{\text{size}(\scriptscriptstyle{\bar{\mathbb{P}}_0^{Q_j}})} \begin{aligned}
 % &\left.\begin{matrix}
%  \partialphantom m = 1  \\[0.5em]
 % \partialphantom m = 2  \\[0.5em]
%  \cdots \\[0.5em]
%  \partialphantom m = M  \\[0.5em]
%  \end{matrix} \right\} %
%  p = 1\\
  %&\begin{matrix}
 % \\[-1.67em]\phantom{\cdots}\cdots
 % \end{matrix}\\ %
 &\left.
 \begin{matrix}
  %\partialphantom m = 2  \\[0.5em]
 \hspace{-2cm} \\[0.75em] 
  %\cdots \\[0.5em]  
  \end{matrix} \hspace{-.9cm}\right\}
 \scriptstyle{ \bar{\mathbb{P}}_0^{Q_j}(l\,:\,m,\,:)}\\ 
 \end{aligned} 
 \end{align}     
 where $\bar{\mathbb{P}}_0^{Q_j}(:,l:m)$ and $\bar{\mathbb{P}}_0^{Q_j}(l:m,:)$, respectively, denote $l$-th to $m$-th columns and rows of $\bar{\mathbb{P}}_0^{Q_j}$.

The values of the major agent's continuous state before and after switching at $t_1$ satisfy the following jump map 
\begin{equation}\label{JumpTranMap1}
 x_{0}^{ex,Q_{1}}(t_{1})=\Psi_{0,\sigma_1}x_{0}^{ex,Q_{0}}(t_{1}-).
 \end{equation}
For the transition between $Q_0$ and case (i) for $Q_1$ where $Q_1 = q_{{}_0^2ab}$ the map $\Psi_{0,\sigma_1}$ is the identity matrix, i.e.
\begin{equation}\label{MajorTransition1ab2ab}
\Psi_{0,\sigma_1} = \Psi_{0,\sigma_{q_{{}_0^1ab},q_{{}_0^2ab}}} = I_{3n}.
 \end{equation} 
This transition is not accompanied by change in the dimension of the major agent's extended state. Furthermore, the weight matrix for the corresponding switching cost is
\begin{equation}
\mathbb{C}_{0,\sigma_1} = \mathbb{C}_{0,\sigma_{q_{{}_0^1ab},q_{{}_0^2ab}}} = 0_{3n \times 3n}.
\end{equation}
For the transition between $Q_0$ and case (ii) where $Q_1 = q_{{}_0^1a}$
 \begin{gather} \label{MajorTransition1ab1a}
 \Psi_{0,\sigma_{1}} = \Psi_{0,\sigma_{q_{{}_0^1ab},q_{{}_0^1a}}} = \left[ \begin{array}{ccc} 
 I_n & 0_{n \times n} & 0_{n \times n}\\
 0_{n \times n} & I_n & 0_{n \times n} \end{array} \right],\allowdisplaybreaks\\
\mathbb{C}_{0,\sigma_1} = \mathbb{C}_{0,\sigma_{q_{{}_0^1ab},q_{{}_0^1a}}} = M_{0}^{q_{{}_0^1ab}}(2n+1:3n). 
%= 
%\left[\begin{array}{ccc}
%0_{n \times n} & 0_{n \times n} & \bar{\mathbb{P}}_{0,13}^{Q_0}\\
%0_{n \times n} & 0_{n \times n} & \bar{\mathbb{P}}_{0,23}^{Q_0}\\
%\bar{\mathbb{P}}_{0,31}^{Q_0} & \bar{\mathbb{P}}_{0,32}^{Q_0} &
%\bar{\mathbb{P}}_{0,33}^{Q_0}
%\end{array}\right],
\end{gather}
%where $\{\bar{\mathbb{P}}_{0,ij}^{Q_0},\, i,j= 1,2,3 \}$ are defined in \eqref{Q0SCost}. 
For the transition between $Q_0$ and case (iii) where $Q_1 = q_{{}_0^1b}$
 \begin{align} \label{MajorTransition1ab1b}
 \Psi_{0,\sigma_1} = \Psi_{0,\sigma_{q_{{}_0^1ab},q_{{}_0^1b}}} = \left[ \begin{array}{ccc} 
 I_n & 0_{n \times n} & 0_{n \times n}\\
 0_{n \times n} & 0_{n \times n} & I_n \end{array} \right],\allowdisplaybreaks \\
\mathbb{C}_{0,\sigma_1} = \mathbb{C}_{0,\sigma_{q_{{}_0^1ab},q_{{}_0^1b}}} = M_{0}^{q_{{}_0^1ab}}(n+1:2n). %\left[\begin{array}{ccc}
%0_{n \times n} & \bar{\mathbb{P}}_{0,12}^{Q_0} & 0_{n \times n}\\
%\bar{\mathbb{P}}_{0,21}^{Q_0} &
%\bar{\mathbb{P}}_{0,22}^{Q_0} & 
%\bar{\mathbb{P}}_{0,23}^{Q_0}\\
%0_{n \times n} &
%\bar{\mathbb{P}}_{0,32}^{Q_0} &
%0_{n \times n}
%\end{array}\right],
\end{align}
%where $\{\bar{\mathbb{P}}_{0,ij}^{Q_0},\, i,j= 1,2,3 \}$ are defined in \eqref{Q0SCost}. 
The values of the major agent's continuous state before and after the switching at $t_2$ satisfy the jump transition map 
\begin{equation}\label{JumpTranMap2}
 x_{0}^{ex,Q_{2}}(t_{2})=\Psi_{0,\sigma_2}x_{0}^{ex,Q_{1}}(t_{2}-),
 % \vspace{-0.5cm}
 \end{equation}
% \vspace{-0.3cm}
where 
\begin{equation}
\Psi_{0,\sigma_2} = \begin{cases}
\Psi_{0,\sigma_{q_{{}_0^1a},q_{{}_0^2a}}}= I_{2n}, \\ %&\text{for $Q_1 =q_{{}_0^1a}  \rightarrow Q_2 = q_{{}_0^2a}$,} \\
\Psi_{0,\sigma_{q_{{}_0^1a},q_{{}_0^1}}}= \begin{bmatrix} I_n & 0_{n \times n}\end{bmatrix}, \\%&\text{for $Q_1 =q_{{}_0^1a} \rightarrow Q_2 = q_{{}_0^1}$,}\\
\Psi_{0,\sigma_{q_{{}_0^2ab},q_{{}_0^2a}}}= \begin{bmatrix}
I_n & 0_{n \times n} & 0_{n \times n} \\
0_{n \times n} & I_n & 0_{n \times n}
\end{bmatrix},\\%&\text{for $Q_1 =q_{{}_0^2ab} \rightarrow Q_2 = q_{{}_0^2a}$,}\vspace{0.15cm}\\
\Psi_{0,\sigma_{q_{{}_0^2ab},q_{{}_0^2b}}}= \begin{bmatrix}
 I_n & 0_{n \times n} & 0_{n \times n}\\
 0_{n \times n} & 0_{n \times n} & I_n 
 \end{bmatrix},\\%& \text{for $Q_1 =q_{{}_0^2ab}  \rightarrow Q_2 = q_{{}_0^2b}$,}\\
\Psi_{0,\sigma_{q_{{}_0^1b},q_{{}_0^2b}}}= I_{2n \times 2n},\\%&\text{for $Q_1 =q_{{}_0^1b} \rightarrow Q_2 = q_{{}_0^2b}$,}\\
\Psi_{0,\sigma_{q_{{}_0^1b},q_{{}_0^1}}}= \begin{bmatrix}
I_n & 0_{n \times n}
\end{bmatrix}, %&\text{for $Q_1 =q_{{}_0^1b} \rightarrow Q_2 = q_{{}_0^1}$.}
\end{cases}
\end{equation}
Furthermore, the matrix coefficient $\mathbb{C}_{0,\sigma_2}$ of the switching cost at $t_2$ for each case is defined as
\begin{equation}
\mathbb{C}_{0,\sigma_2} = \begin{cases}
\mathbb{C}_{0,\sigma_{q_{{}_0^1a},q_{{}_0^2a}}}= 0_{2n \times 2n},\\  %&\text{for $Q_1 =q_{{}_0^1a} \rightarrow Q_2 = q_{{}_0^2a}$,} \\\allowdisplaybreaks
\mathbb{C}_{0,\sigma_{q_{{}_0^1a},q_{{}_0^1}}}= M_{0}^{q_{{}_0^1a}}(n+1:2n),\\%&\hspace{0.5cm}\text{for transition from $Q_1 =q_{{}_0^1a} $ to $Q_2 = q_{{}_0^1}$,}\allowdisplaybreaks\\%=\left[\begin{array}{cc}
%0_{n \times n} & \bar{\mathbb{P}}_{0,12}^{q_{{}^1_0a}} \\
%\bar{\mathbb{P}}_{0,21}^{q_{{}^1_0a}} & \bar{\mathbb{P}}_{0,22}^{q_{{}^1_0a}} 
%\end{array}\right],
\mathbb{C}_{0,\sigma_{q_{{}_0^2ab},q_{{}_0^2a}}}=  M_{0}^{q_{{}_0^2ab}}(2n+1:3n),\\ % &\hspace{0.5cm}\text{for transition from $Q_1=q_{{}_0^2ab} $ to $Q_2 = q_{{}_0^2a}$,}\vspace{0.15cm}\allowdisplaybreaks\\

%\left[\begin{array}{ccc}
%0_{n \times n} & 0_{n \times n} & \bar{\mathbb{P}}_{0,13}^{q_{{}^2_0ab}}\\
%0_{n \times n} & 0_{n \times n} & 
%\bar{\mathbb{P}}_{0,23}^{q_{{}^2_0ab}}\\
%\bar{\mathbb{P}}_{0,31}^{q_{{}^2_0ab}} &
%\bar{\mathbb{P}}_{0,32}^{q_{{}^2_0ab}} &
%\bar{\mathbb{P}}_{0,33}^{q_{{}^2_0ab}}
%\end{array}\right],

\mathbb{C}_{0,\sigma_{q_{{}_0^2ab},q_{{}_0^2b}}}= M_{0}^{q_{{}_0^2ab}}(n+1:2n),\\   %&\hspace{0.5cm}\text{for transition from $Q_1 =q_{{}_0^2ab} $ to $Q_2 = q_{{}_0^2b}$,}\allowdisplaybreaks\\

%\left[\begin{array}{ccc}
%0_{n \times n} & \bar{\mathbb{P}}_{0,12}^{q_{{}^2_0ab}} & 0_{n \times n}\\
%\bar{\mathbb{P}}_{0,21}^{q_{{}^2_0ab}} &
%\bar{\mathbb{P}}_{0,22}^{q_{{}^2_0ab}} & 
%\bar{\mathbb{P}}_{0,23}^{q_{{}^2_0ab}}\\
%0_{n \times n} &
%\bar{\mathbb{P}}_{0,32}^{q_{{}^2_0ab}} & 0_{n \times n} 
%\end{array}\right],

\mathbb{C}_{0,\sigma_{q_{{}_0^1b},q_{{}_0^2b}}}= 0_{2n \times 2n},\\%&\hspace{0.5cm}\text{for transition from $Q_1 =q_{{}_0^1b}$ to $Q_2 = q_{{}_0^2b}$,}\allowdisplaybreaks\\
\mathbb{C}_{0,q_{{}_0^1b},q_{{}_0^1}}= M_{0}^{q_{{}_0^1b}}(n+1:2n). %\\&\hspace{0.5cm}\text{for transition from $Q_1 =q_{{}_0^1b}$ to $Q_2 = q_{{}_0^1}$.}
%\left[\begin{array}{cc}
%0_{n \times n} & \bar{\mathbb{P}}_{0,12}^{q_{{}^1_0b}} \\
%\bar{\mathbb{P}}_{0,21}^{q_{{}^1_0b}} & \bar{\mathbb{P}}_{0,22}^{q_{{}^1_0b}} %\end{array}\right], 
\end{cases} \label{C02SCost}
\end{equation}
%where the corresponding entries in \eqref{C02SCost} are defined in \eqref{q02abSCost},  \eqref{q01aSCost}, and \eqref{q01bSCost}.
The values of the major agent's continuous state before and after the switching at $t_3$ satisfy the following jump map 
\begin{gather}
 x_{0}^{ex,Q_{3}}(t_{3})=\Psi_{0,\sigma_3}x_{0}^{ex,Q_{2}}(t_{3}-),\label{JumpTranMap3}\\
 \Psi_{0,\sigma_3} = \begin{cases}
\Psi_{0,\sigma_{q_{{}_0^2a},q_{{}_0^2}}} = \begin{bmatrix}
I_n & 0_{n \times n} \end{bmatrix}, \\%&\hspace{0.5cm}\text{for transition from $Q_2 =q_{{}_0^2a} $ to $Q_3$,}\\
\Psi_{0,\sigma_{q_{{}_0^1},q_{{}_0^2}}} = I_n, \\%&\hspace{0.5cm}\text{for transition from $Q_2 =q_{{}_0^1}$ to $Q_3$,}\\
\Psi_{0,\sigma_{q_{{}_0^2b},q_{{}_0^2}}} = \left[ \begin{array}{cc}
I_n & 0_{n \times n} \end{array} \right].\label{eq:Psi03}%&\hspace{0.5cm}\text{for transition from $Q_1=q_{{}_0^2b}$ to $Q_3$.}\\
\end{cases}
 \end{gather}
% where 
% \begin{align}\label{eq:Psi03}
% \Psi_{0,\sigma_3} = \begin{cases}
% \Psi_{0,\sigma_{q_{{}_0^2a},q_{{}_0^2}}} = \begin{bmatrix}
% I_n & 0_{n \times n} \end{bmatrix}, \\%&\hspace{0.5cm}\text{for transition from $Q_2 =q_{{}_0^2a} $ to $Q_3$,}\\
% \Psi_{0,\sigma_{q_{{}_0^1},q_{{}_0^2}}} = I_n, \\%&\hspace{0.5cm}\text{for transition from $Q_2 =q_{{}_0^1}$ to $Q_3$,}\\
% \Psi_{0,\sigma_{q_{{}_0^2b},q_{{}_0^2}}} = \left[ \begin{array}{cc}
% I_n & 0_{n \times n} \end{array} \right].%&\hspace{0.5cm}\text{for transition from $Q_1=q_{{}_0^2b}$ to $Q_3$.}\\
% \end{cases}
% \end{align}
Accordingly, the matrix coefficient $\mathbb{C}_{0,3}$ of the switching cost at $t_3$ for each case is given by
\begin{align}
\mathbb{C}_{0,\sigma_3} = \begin{cases}
\mathbb{C}_{0,\sigma_{q_{{}_0^2a},q_{{}_0^2}}} = M_{0}^{q_{{}_0^2a}}(n+1:2n),\\%&\hspace{0.5cm}\text{for transition from $Q_2 =q_{{}_0^2a} $ to $Q_3$,}\\
%\left[\begin{array}{cc}
%0_{n \times n} & \bar{\mathbb{P}}_{0,12}^{q_{{}^2_0a}} \\
%\bar{\mathbb{P}}_{0,21}^{q_{{}^2_0a}} & \bar{\mathbb{P}}_{0,22}^{q_{{}^2_0a}} 
%\end{array}\right], 
\mathbb{C}_{0,\sigma_{q_{{}_0^1},q_{{}_0^2}}} = 0_{n \times n}, \\%&\hspace{0.5cm}\text{for transition from $Q_2 =q_{{}_0^1}$ to $Q_3$,}\\
\mathbb{C}_{0,\sigma_{q_{{}_0^2b},q_{{}_0^2}}} = M_{0}^{q_{{}_0^2b}}(n+1:2n).% &\hspace{0.5cm}\text{for transition from $Q_1=q_{{}_0^2b}$ to $Q_3$,}\\
%\left[\begin{array}{cc}
%0_{n \times n} & \bar{\mathbb{P}}_{0,12}^{q_{{}^2_0b}} \\
%\bar{\mathbb{P}}_{0,21}^{q_{{}^2_0b}} & \bar{\mathbb{P}}_{0,22}^{q_{{}^2_0b}} 
%\end{array}\right],
\end{cases} \label{C03SCost}
\end{align}
Notice that some of the transitions of \eqref{JumpTranMap1}, \eqref{JumpTranMap2}, \eqref{JumpTranMap3} are between the spaces of the same dimension such as \eqref{MajorTransition1ab2ab} while other transitions may be accompanied by changes in the dimension of the state space, e.g. \eqref{MajorTransition1ab1a} is a mapping from $\mathbb{R}^{3n}$ into $\mathbb{R}^{2n}$. These dimension changes are permitted in the stochastic hybrid systems framework of \cite{APPECCDC2016,APPEC2017IFAC} (see \cite{APPEC2016NAHS} for another motivating example for change of dimension at switching).
\subsubsection{Best Response Hybrid Control Action}\label{sec:BRHybCA-major}
To obtain the best response hybrid control action for the major agent in the infinite population, we utilize \textit{Theorem} \ref{thm:HybridLQG} in Section \ref{sec:HybridLQG} developed for single agent hybrid LQG optimal control problems.
By the definition of the terms $\mathbb{D}_0^{Q_j}$, they automatically satisfy the  condition \eqref{A1EquivalentDiffusion} (see Section \ref{sec:HybridLQG}), or equivalently condition A1 in \cite[Eq. (3)]{APPECCDC2016} as
\begin{align} \label{DiffusionSwitchingRelation}
\mathbb{D}_0^{Q_j} = \Psi_{0,\sigma_j} \mathbb{D}_0^{Q_{j-1}}, \hspace{20pt} j = 1,2,3,
\end{align}
holds for all the jump transition maps introduced in this section. Moreover, it is assumed conditions \eqref{DeltaHzero}-\eqref{DeltaHnegative} in Section \ref{sec:HybridLQG} hold. Therefore, the optimal controlled switching times for the major agent are $\mathcal{F}_t$-independent. Then an application of the hybrid LQG optimal control theory (see \textit{Theorem}~\ref{thm:HybridLQG}) yields the infinite population best response hybrid control action for discrete states $\{Q_0,\dots,Q_3\}$ as in 
\begin{align}
u_0^{Q_j, \circ} (t) &= - [R_{0}^{Q_j}]^{-1} [\mathbb{B}_{0}^{Q_j}]^T \Pi_{0}^{Q_j} (t) \, x^{ex,Q_j}_0 (t),\label{majorControl_hybrid}\\
-\dot{\Pi}_0^{Q_j} &= \Pi_0^{Q_j} \mathbb{A}_0^{Q_j} + [\mathbb{A}_{0}^{Q_j}]^T \Pi_0^{Q_j}\nonumber \allowdisplaybreaks\\
&\hspace{1cm}- \Pi_0^{Q_j} \mathbb{B}_0^{Q_j} [R_{0}^{Q_j}]^{-1} [\mathbb{B}_{0}^{Q_j}]^T \Pi_0^{Q_j} + \mathbb{P}_0^{Q_j},\label{RiccatiMajor}
 \end{align}
 subject to the terminal and boundary conditions
 \begin{gather}
 \Pi_0^{Q_3}(T) = \bar{\mathbb{P}}_0^{Q_3} ,\label{MajorTerminalC}\\[-2pt]
 \Pi^{Q_{j-1}}_{0}(t_j) = \Psi_{0,\sigma_j}^T \Pi^{Q_{j}}_{0}(t_j) \Psi_{0,\sigma_j} + \mathbb{C}_{0,\sigma_j} ,\label{MajorBoundaryC}
\end{gather}
 \vspace{-9mm}
\begin{multline}\label{majorHcontinuity_hybrid}
\mathbb{P}_0^{Q_{j-1}} +  \Pi^{Q_{j-1}}_{0}(t_j) \mathbb{A}_0^{Q_{j-1}} +[\mathbb{A}_{0}^{Q_{j-1}}]^T  \Pi^{Q_{j-1}}_{0}(t_j) \allowdisplaybreaks\\-  \Pi^{Q_{j-1}}_{0}(t_j) \mathbb{B}_0^{Q_{j-1}} [R_{0}^{Q_{j-1}}]^{-1}[\mathbb{B}_{0}^ {Q_{j-1}}]^T  \Pi^{Q_{j-1}}_{0}(t_j) \allowdisplaybreaks\\= \Psi_{0,\sigma_j}^T \Big( \mathbb{P}_0^{Q_j} + \Pi_{0}^{Q_j}(t_j) \mathbb{A}_{0}^{Q_j} + [\mathbb{A}_{0}^{Q_j}]^T \Pi_{0}^{Q_j}(t_j) \allowdisplaybreaks\\-\Pi_{0}^{Q_j}(t_j)\mathbb{B}_{0}^{Q_j} [R_{0}^{Q_j}]^{-1} [\mathbb{B}_{0}^{Q_j}]^T \Pi_0^{Q_j}(t_j) \Big) \Psi_{0,\sigma_j} + \frac{\partial \mathbb{C}_{0,\sigma_j}}{\partial t}\Big |_{t=t_j},
\end{multline}
where $t_j$, $j=1, 2, 3$, indicate the times of change in the system due to the major agent's switching of dynamics or cessation of subpopulations of minor agents.
\subsection{Minor Agents: Infinite Population} \label{sec:MinorInf}
%\vspace{-6pt}
%For brevity, the notation $(.)_{a/l}$ is used in the rest of this paper to denote the matrices and parameters corresponding to a generic acquirer or a liquidator agent, respectively.  
\begingroup
\subsubsection{Hybrid Dynamics and Costs}\label{sec:hybDynCost-minor}
The extended dynamics for a generic minor agent $\mathcal{A}_i^k, \, 1\leq i \leq N$, in the population $k\triangleq a,b,$ with the extended state $x^{ex, Q_j}_i$ has a general form as in 
\begin{equation} 
dx_{i}^{ex,Q_{j}}=(\mathbb{A}_{k}^{Q_{j}}x_{i}^{ex,Q_{j}} +\mathbb{M}_{k}^{Q_{j}}+\mathbb{B}_{k}^{Q_{j}}u_{i}^{Q_{i}})dt+\mathbb{D}_{k}^{Q_{j}}dW_{i}^{Q_j},  
\label{MinorAcquirerHybridDynamics_hybrid}
\end{equation}
where 
\begin{gather}
\mathbb{A}_{k}^{Q_{j}}\!=\!\begin{bmatrix}
A_{k} & \begin{bmatrix} G_{k} &\pi^{Q_j} \otimes F_{k} \end{bmatrix}\\
0_{\bullet \times \bullet} & \mathbb{A}_{0}^{Q_{j}}-\mathbb{B}_{0}^{Q_{j}}R_{0,Q_{j}}^{-1}\mathbb{B}_{0,Q_{j}}^{T}\Pi_{0}^{Q_{j}}
\end{bmatrix},\, \,
\mathbb{M}_{k}^{Q_{j}}\!\!=\!\!\begin{bmatrix}
0_{n\times1}\\
\mathbb{M}_{0}^{Q_j}
\end{bmatrix},\nonumber \allowdisplaybreaks\\
\mathbb{B}_{k}^{Q_{j}}=\begin{bmatrix}
B_{k}\\
0_{\bullet \times \bullet}
\end{bmatrix}, ~
\mathbb{D}_{k}^{Q_{j}}=\begin{bmatrix}
D_{k} & 0_{\bullet \times \bullet}\\
0_{\bullet \times \bullet} & \mathbb{D}_{0}^{Q_{j}}
\end{bmatrix}, ~ 
W_i^{Q_j} = \begin{bmatrix}
w_i\\
w_0^{Q_j}
\end{bmatrix}. \label{minorExtDynHybMatrix}
\end{gather}
Notice that in \eqref{MinorAcquirerHybridDynamics_hybrid} the major agent's closed-loop dynamics at discrete state $Q_j, \, 0\leq j \leq 3$, given by \eqref{MajorHybridDSE_hybrid} is used to derive the extended dynamics for minor agent $\mathcal{A}_i^{k}$ at discrete state $Q_j, \, 0\leq j \leq 3$. Similar to the major agent's case, $0_{\bullet \times \bullet}$ in \eqref{minorExtDynHybMatrix} denotes a zero matrix of appropriate dimension. 

The cost functional for the extended minor agent $\mathcal{A}^k_i$'s hybrid system is given by
\begin{multline}
J_i^{\infty, k}(u_i, u_{0})= \hf\mathbb{E} \Big [ \Vert x^{ex,Q_\ell}_i(t_s^i) \Vert^2_{\bar{\mathbb{P}}_k^{Q_\ell}} + \sum_{j=1}^{\ell} \Vert x_i^{ex,Q_j}(t_j^-)\Vert^2_{\mathbb{C}_{i,\sigma_j}^{k}}\\+ \sum_{j=0}^{\ell} \int_{t_j}^{t_{j+1}} \big ( \Vert x^{ex,Q_j}_i(s) \Vert^2_{\mathbb{P}_k^{Q_j}} + \Vert u_i^{Q_j}(s) \Vert^2_{R_k} \big ) ds \Big], \label{minorCostQuadraticHyb}
\end{multline}
%where $t_s^i$, $i \in \{1, 2, 3\}$ denotes the $i^{\text{th}}$ time a switching occurs in the population dynamics and for the case where the switching corresponds to an individual and its corresponding group of agents $\mathcal{A}^k_i$ quits the system at $t_s^i$ with $\ell \in \{1, 2\}$ being the index of the associate group index.
where $Q_{\ell}$ denotes the discrete state after which minor agent $\mathcal{A}^k_i$ quits the system at the individual stopping time $t_s^i$ and \mbox{$\ell \in \{0,1, 2\}$} denotes the index of the associate discrete state. The weight matrices associated with the terminal cost (first term) and the running cost (third term) in \eqref{minorCostQuadraticHyb} are, respectively, given by
\begin{gather}
\bar{\mathbb{P}}_k^{Q_\ell} = \bar{P}_k,\nonumber\\
\mathbb{P}_k^{Q_j} =  [I_{n}, -H_1^k, -\pi^{Q_j}\otimes H_2^k]^T P_k [I_{n}, -H_1^k, -\pi^{Q_j}\otimes H_2^k],\nonumber\\
\bar{\mathbb{P}}_k^{Q_j} =  [I_{n}, -H_1^k, -\pi^{Q_j}\otimes H_2^k]^T \bar{P}_k [I_{n}, -H_1^k, -\pi^{Q_j}\otimes H_2^k],\label{minorExMatGenCost}
\end{gather}
where $\bar{\mathbb{P}}_k^{Q_j}$ shall be used to specify the weight matrix $\mathbb{C}^k_{i,\sigma_j}$ associated with the switching cost (second term) in \eqref{minorCostQuadraticHyb} in a similar manner to that of the major agent in Section~\ref{sec:majorJumpMapNswitchingCost}. The values of minor agent $\mathcal{A}_i^k$ continuous state before and after the switching time $t_j$ satisfy the following jump transition map 
\begin{align}\label{JumpTranMinor}
 x_{i}^{ex,Q_{j}}(t_{j})=\Psi_{i,\sigma_j}^k x_{i}^{ex,Q_{j-1}}(t_{j}-).
 \end{align}
 The realizations of $\mathbb{C}^k_{i,\sigma_j}$ and $\Psi_{i,\sigma_j}^k$ associated with the switching times $t_j, j = 1,2,3$, are presented in Appendix \ref{sec:minorJumpMapNSwitchingCost}.
\subsubsection{Best Response Hybrid Control Actions}\label{sec:BRHybCA-minor}
The optimal stopping problem for a minor agent is equivalent to a hybrid optimal control problem in which the dynamics and costs become zero after stopping. Let us assume that minor agent $\mathcal{A}_i^k$ stops at time $t_s^k$ after the discrete state $Q_{\ell}, \, \ell \in \{ 0,1, 2  \}$. The definitions for $\mathbb{D}_k^{Q_j}$ directly result in the satisfaction of condition \eqref{A1EquivalentDiffusion} (see Section \ref{sec:HybridLQG}), or equivalently condition A1 in \cite[Eq. (3)]{APPECCDC2016}, i.e. 
\begin{align}
\mathbb{D}_k^{Q_j} = \Psi_{i,\sigma_j}^k \mathbb{D}_k^{Q_{j-1}}, \quad j \in \{1,\dots,\ell\}, \quad k \triangleq a,b.
\end{align}
Furthermore, it is assumed in this paper that conditions \eqref{DeltaHzero}-\eqref{DeltaHnegative}, and \eqref{stopHzero}-\eqref{stopHnegative}, respectively, hold on the parameters $\mathbb{A}_{k},\mathbb{B}_{k},\mathbb{D}_{k}, \mathbb{\bar{P}}_{k}, \mathbb{P}_{k}, \Psi^{k}_{i,\sigma}$, etc., of the extended system and, consequently, the optimal stopping times $t_s^k$ and switching times $t_j$ %(whenever it is the case)
exist and are determined by the solutions to \eqref{DeltaHzero}-\eqref{DeltaHnegative}, and \eqref{stopHzero}-\eqref{stopHnegative}. Moreover, the optimal stopping and switching times  %\sout{. Hence, the optimal stopping time for each minor agent is $\mathcal{F}_t$-independent and only depends on its dynamical parameters which} 
are $\mathcal{F}_t$-independent and depend only  on the dynamical parameters; this implies that all minor agents of the same type switch and stop at the same instant. Then the application of the results of \textit{Theorem \ref{thm:HybridLQG}} and \textit{Corollary \ref{thm:HybridLQGstop}} yields the infinite population best response strategies for the discrete states $\{Q_0,\dots,Q_\ell \}$ given by 
\begin{align}\label{minorControl}
u_{i}^{Q_j, \circ} (t)= - R_k^{-1} [\mathbb{B}_{k}^{Q_j}]^T \Pi_{k}^{Q_j} (t) \, x^{ex,Q_j}_i (t),
\end{align}
with
\begin{equation}\label{minorRic_HybridMFG}
-\dot{\Pi}_k^{Q_j} = \Pi_k^{Q_j} \mathbb{A}_k^{Q_j} + \mathbb{A}_{k,Q_j}^T \Pi_k^{Q_j} - \Pi_k^{Q_j} \mathbb{B}_k^{Q_j} R_{k}^{-1} [\mathbb{B}_{k}^{Q_j}]^T \Pi_k^{Q_j} + \mathbb{P}_k^{Q_j}
 \end{equation}
subject to the terminal conditions 
 \begin{align}
 \Pi_k^{Q_\ell}(t_s^k) &= \bar{\mathbb{P}}_k^{Q_\ell},\allowdisplaybreaks\\
  \Big(\mathbb{P}_{k}^{Q_{\ell}}+\bar{\mathbb{P}}_{k}^{Q_{\ell}}\mathbb{A}_{k}^{Q_{\ell}}+\mathbb{A}_{k,Q_{\ell}}^{T}\bar{\mathbb{P}}_{k}^{Q_{\ell}}
&-\bar{\mathbb{P}}_{k}^{Q_{\ell}}\mathbb{B}_{k}^{Q_{\ell}}R_{k}^{-1}[\mathbb{B}_{k}^{Q_{\ell}}]^{T}\bar{\mathbb{P}}_{k}^{Q_{\ell}} \Big)_{t=t_{s}^k} \nonumber\\&\hspace{0.5cm}=\frac{\partial \mathbb{C}^k_{i,\sigma_\ell}}{\partial t}\Big |_{t=t_s^k}, \label{eq:minorStoppCond}
\end{align}
and the boundary conditions
\begin{align}
\Pi_{k}^{Q_{j-1}}(t_{j})=& {[\Psi_{i,\sigma_j}^{k}]}^T\Pi_{k}^{Q_{j}}(t_{j})\Psi_{i,\sigma_j}^k + \mb{C}_{i,\sigma_j}^k,\label{minorRicJump}
\end{align}
\begin{align}
\mathbb{P}_{k}^{Q_{j-1}}+\Pi_{k}^{Q_{j-1}}(t_{j})&\mathbb{A}_{k}^{Q_{j-1}}+[\mathbb{A}_{k}^{Q_{j-1}}]^{T}\Pi_{k}^{Q_{j-1}}(t_{j}) \nonumber\allowdisplaybreaks\\-\Pi_{k}^{Q_{j-1}}(t_{j})&\mathbb{B}_{k}^{Q_{j-1}}R_{k}^{-1}[\mathbb{B}_{k}^{Q_{j-1}}]^{T}\Pi_{k}^{Q_{j-1}}(t_{j})\nonumber\allowdisplaybreaks\\
={[\Psi_{i,\sigma_j}^{k}]}^T\Big(\mathbb{P}_{k}^{Q_{j}}&+\Pi_{k}^{Q_{j}}(t_{j})\mathbb{A}_{k}^{Q_{j}}+[\mathbb{A}_{k}^{Q_{j}}]^{T}\Pi_{k}^{Q_{j}}(t_{j})\nonumber\allowdisplaybreaks\\-\Pi_{k}^{Q_{j}}(t_{j})\mathbb{B}_{k}^{Q_{j}}R_{k}^{-1}&[\mathbb{B}_{k}^{Q_{j}}]^{T}\Pi_{k}^{Q_{j}}(t_{j})\Big)\Psi_{i,\sigma_j}^k + \frac{\partial \mathbb{C}_{i,\sigma_j}^k}{\partial t}\Big|_{t=t_j},\label{MinorHcontinuity}
\end{align}
if $t_j < t_s^k$, where $\{t_j$, $j \in \{1,\dots,\ell\}\}$ indicate the times of change in the system due to the major agent's switching of dynamics or cessation of the other subpopulation of minor agents. %\ali{I think this last sentence should be removed, particularly since (82) and (83) are the terminal conditions for (81).} \dena{revised.} 
We observe that for the case where the subpopulation $k, \, k\triangleq a, b$, stops at time $t_1$, the boundary conditions \eqref{minorRicJump} and \eqref{MinorHcontinuity} are irrelevant for the agents belonging to the quitting subpopulation.
\subsection{Hybrid Mean Field Consistency Equations and $\epsilon$-Nash Equilibrium} \label{sec:ConsistencyCondition}% \vspace{-8pt}
Let us define 
\begin{align}
\Pi_k^{Q_j} &= \left[ \begin{array}{ccc}
\Pi_{k,11}^{Q_j} & \Pi_{k,12}^{Q_j} & \Pi_{k,13}^{Q_j} \\
\Pi_{k,21}^{Q_j} & \Pi_{k,22}^{Q_j} & \Pi_{k,23}^{Q_j}\\
\Pi_{k,31}^{Q_j} & \Pi_{k,32}^{Q_j} & \Pi_{k,33}^{Q_j}
\end{array} \right] ,\quad k \triangleq a,b, \nonumber\\
\mathbf{e}_k^{Q_j} =& \begin{cases}
  I_n  \quad &\mbox{if} \, \, \, \bar{x}^{Q_j} = \bar{x}_k,\\
 [ I_n,0_{n \times n}] \quad &\mbox{if} \, \, \{\bar{x}^{Q_j} \neq \bar{x}_k\} \wedge\{ k=a \},\\
 [0_{n \times n}, I_n] \quad &\mbox{if} \, \,\{\bar{x}^{Q_j} \neq \bar{x}_k\} \wedge \{k=b\}.
 \end{cases}
\end{align}
%where $I_n$ is an $n \times n$ identity matrix. 
We substitute the set of obtained best response strategies \eqref{minorControl} in \eqref{MinorStandardDynamics_hybrid}. Then we take an average over the corresponding subpopulation and take its limit as $N_t \rightarrow \infty$. In order to generate a mean field game equilibrium the obtained equation and \eqref{MFequation_hybrid} must correspond to the same dynamical system generating the mean field. Consequently we obtain the Consistency Condition equations, determining \iffalse the components of $\bar{A}$, $\bar{G}$, $\bar{H}$, $\bar{L}$, $\bar{J}$, and $\bar{m}$ in \eqref{MeanFieldEq}, given by the following compact set of equations
Then, by consistency requirement, the set of obtained best response strategies \eqref{minorControl} when substituted in \eqref{MinorAcqStandardDynamics_hybrid} or \eqref{MinorLiqStandardDynamics_hybrid} and averaged over the subpopulation should converge, as $N_t \rightarrow \infty$, to the mean field equation \eqref{MFequation_hybrid} which was used to derive the strategies \cite{Huang2010, TAC17arXiv}. Accordingly, a compact description of the hybrid major minor mean field equations determining\fi $\bar{A}^{Q_j}$, $\bar{G}^{Q_j}$, $\bar{m}^{Q_j}$ given by
\begin{align}
&\dot{\Pi}_{0}^{Q_{j}}=\Pi_{0}^{Q_{j}}\mathbb{B}_{0}^{Q_{j}}R_{0}^{-1}[\mathbb{B}_{0}^{Q_{j}}]^{T}\Pi_{0}^{Q_{j}}-\Pi_{0}^{Q_{j}}\mathbb{A}_{0}^{Q_{j}}-[\mathbb{A}_{0}^{Q_{j}}]^{T}\Pi_{0}^{Q_{j}}-\mathbb{P}_{0}^{Q_{j}}, \nonumber \allowdisplaybreaks \\ 
&\dot{\Pi}_{k}^{Q_{j}}=\Pi_{k}^{Q_{j}}\mathbb{B}_{k}^{Q_{j}}R_{k}^{-1}[\mathbb{B}_{k}^{Q_{j}}]^{T}\Pi_{k}^{Q_{j}}-\Pi_{k}^{Q_{j}}\mathbb{A}_{k}^{Q_{j}}-[\mathbb{A}_{k}^{Q_{j}}]^{T}\Pi_{k}^{Q_{j}}-\mathbb{P}_{k}^{Q_{j}},  \nonumber \allowdisplaybreaks \\ 
&\bar{A}^{Q_j}_k = [A_k - B_k R_{k}^{-1} B_k^T \Pi_{k,11}^{Q_j}] \mathbf{e}_k^{Q_j} + \pi^{Q_j} \otimes F_k - B_k R_k^{-1} B_k^T \Pi_{k,13}^{Q_j},  \nonumber \allowdisplaybreaks\\ 
&\bar{G}^{Q_j}_k  = G_k  - B_k R_k^{-1} B_k^T \Pi_{k,12}^{Q_j}, \nonumber \allowdisplaybreaks \\ 
&\bar{m}^{Q_j}_k = 0,~~ \label{mfeqConsistency_hybrid}\allowdisplaybreaks
\end{align}   
for each discrete state $Q_j, \, 0 \leq j \leq 3$, and the corresponding population $k,\, k\triangleq a,b$. The set of equations \eqref{mfeqConsistency_hybrid} forms a fixed-point problem for each discrete state $Q_j, 1\leq j \leq 3$, that should be solved by each minor agent in order to compute the matrices in the mean field dynamics.
\begin{assum} \label{MFEquationSolAss}
The parameters in \eqref{MajorStandardDynamics_hybrid}-\eqref{minorCost_hybrid} belong to a non-empty set which yields the existence and uniqueness of the solutions ($\Pi_0^{Q_{j}}$, $\Pi_k^{Q_{j}}$, $\bar{A}_k^{Q_{j}}$, ${\bar{G}}^{Q_j}_k$, ${\bar{m}_k}^{Q_{j}}$) to the resulting set of mean field fixed-point equations \eqref{mfeqConsistency_hybrid} at each discrete state $Q^j,\, 0 \leq j \leq 3$.
\end{assum}
%\begin{assumption} \label{MFEquationSolAss_HybridMFG}
% There exists a stabilizing solution $\Pi_0^{Q_j}$, $\Pi_k^{Q_j}$, $\bar{A}_{k}^{Q_j}$, $\bar{G}_{k}^{Q_j}$, $1\leq j\leq 3,\, k\triangleq a,b$, to the major-minor mean field equations (\ref{mfeqConsistency_hybrid}) in the sense that the matrices 
%   \begin{gather*}
%  \mathbb{A}_0^{Q_j}-\mathbb{B}_0^{Q_j} [R_0^{Q_j}]^{-1}[\mathbb{B}_0^{Q_j}]^T \Pi_0^{Q_j}, \\
%    \mathbb{A}_{k}^{Q_j} -\mathbb{B}_{k}^{Q_j} R_k^{-1} [\mathbb{B}_{k}^{Q_j}]^T \Pi_{k}^{Q_j},
%  \end{gather*}
%  are asymptotically stable.
%  \end{assumption}
%\begin{assum}\label{ass:detectability_MF}
%The pair $((\mathbb{P}_0^{Q_j})^{\frac{1}{2}}, \mathbb{A}_0^{Q_j})$ is detectable, and for each $k \triangleq a, b$, the pair $((\mathbb{P}_k^{Q_j})^{\frac{1}{2}}, \mathbb{A}_k^{Q_j})$ is detectable.%, where $L_a = Q_0^{1/2}[I, -H_0^\pi]$ and $L_b = Q^{1/2}_k[I, -H_k, -\hat{H}_k^\pi]$.
%\end{assum}
%\begin{assum}\label{ass:stabilizability_MF}
%The pair $(\mathbb{A}_0^{Q_j}, \mathbb{B}_0^{Q_j})$ is stabilizable and $(\mathbb{A}_k^{Q_j}, \mathbb{B}_k^{Q_j})$ is stabilizable for each $k \triangleq a, b$.
%\end{assum}
The following theorem links the infinite population equilibria to the finite population case.  
\begin{thm}[{\small$\epsilon$-Nash Equilibrium for Hybrid LQG MFGs}] \label{thm:ENE_hybrid}
Subject to \textit{Assumptions \ref{ass:major_parameters}-\ref{MFEquationSolAss}}, and subject to the assumption that the equations \eqref{DeltaHzero}-\eqref{DeltaHnegative}, and \eqref{stopHzero}-\eqref{stopHnegative} are satisfied, the $\mc{F}_t$-invariant optimal switching and stopping times $t_1,t_2,t_3$ exist and are 	uniquely determined. In that case system equations \eqref{MajorHybridDSE_hybrid}-\eqref{majorExMatGenCost}, \eqref{MinorAcquirerHybridDynamics_hybrid}-\eqref{minorExMatGenCost}, together with the set of mean field equations \eqref{mfeqConsistency_hybrid} generate a set of control laws  which yields the infinite population Nash equilibrium. When the set of infinite population control laws $\mathcal{U}_{MF}^{N_t} \triangleq \{ {u}_i^{Q_j, \circ}; 0\leq i \leq N_t \},\, 0 \leq j \leq 3,\, 1\leq N_t \leq N < \infty,$ given by \eqref{majorControl_hybrid}, \eqref{minorControl} is applied to the finite population system \eqref{MajorStandardDynamics_hybrid}, \eqref{MinorStandardDynamics_hybrid}, it %results in the following properties:
%\begin{enumerate}
%\item[(i)] All agent systems $\mathcal{A}_i, ~0 \leq i \leq N$, are second order stable. 
%$e^{-\frac{\rho}{2}t}$discounted second order stable in the sense that\\
%$\sup_{t\geq 0, 0\leq i \leq N} e^{-\frac{\rho}{2}t} \mathbb{E} \Big ( {\Vert \hat{X}_{i|\mathcal{F}^y_i}(t) \Vert}^2 + {\Vert \hat{\bar{X}}_{|\mathcal{F}^y_i}(t)\Vert}^2 \Big) < C$, \\
 %with C independent of N;
% \item[(ii)]   $\mathcal{U}_{MF}^{N_t}, 1\leq N_t < \infty$ 
 yields an $\epsilon$-Nash equilibrium for all $\epsilon$, i.e. for all $\epsilon>0$, there exists $N(\epsilon)$ such that for all $N \geq N(\epsilon)$; 
\begin{multline*}
J_i^{Q_j,N}(u_i^{Q_j,\circ},  u_{-i}^{Q_j, \circ})-\epsilon \leq\inf_{u_i \in\mathcal{U}^{N,L}_g } J_i^{Q_j,N}(u_i, u_{-i}^{Q_j,\circ}) \\\leq  J_i^{Q_j,N}( u_i^{Q_j,\circ}, u_{-i}^{Q_j,\circ}),
\end{multline*}
where $J_i^{Q_j,N}(.,.)$ denotes the finite-population cost functional of the generic agent $\mathcal{A}_i$ at the discrete state $Q_j$.
 \hfill $\square$
%\end{enumerate}
 \end{thm}
\textit{Proof.}
See Apprendix \ref{app:NashProof}.
%Applying the approach of \cite{Huang2010} backwards from $T$ along the optimal realization of the sequence $Q_0,\, Q_1,\, Q_2,\, Q_3$ establishes the existence and uniqueness of the Nash equilibrium and $\epsilon$-Nash equilibrium for the infinite population system and finite population system, respectively.    
\subsection{Hybrid Dynamic Programming Methodology}\label{sec:HDP}
The order of the switching and stopping events $Q_0, Q_1, Q_2, Q_3$, if all of them occur, is assumed to be fixed. As depicted in Figure \ref{fig:HybridAutomata} and explained in Section \ref{sec:DiscreteState}, there are three possible realizations for each of the discrete states $Q_1$ and $Q_2$. The optimal sequence of switching, that is to say the discrete trajectory of the system, is determined via dynamic programming backward propagation. For this purpose, the steps below are followed. 

\textit{Step 1}. \textit{(Solving backwards for transitions from $Q_3$ to $Q_2$)}. Equation \eqref{RiccatiMajor} is solved for $\Pi_0^{Q_3}(t)$ backward in time, subject to the terminal condition \eqref{MajorTerminalC}. Then the values for $\Pi_0^{Q_3}(t)$ are substituted in the right hand side of \eqref{MajorBoundaryC} to obtain $\Pi_0^{Q_2}(t)$ for all three realizations of $\Psi_{0,\sigma_3}$ and $\mathbb{C}_{0,\sigma_3}$ given by \eqref{eq:Psi03} and \eqref{C03SCost}, respectively. Next, we substitute $\Pi_0^{Q_2}(t)$ and the corresponding $\Psi_{0,\sigma_3}$ and $\mathbb{C}_{0,\sigma_3}$ in \eqref{majorHcontinuity_hybrid}. Then the time instant at which \eqref{majorHcontinuity_hybrid} holds determines $t_3$ for the transition from the corresponding realization of $Q_2$ to $Q_3$. The transitions from $Q_2 \triangleq q_{{}^2_0b}$  to $Q_3$ or from $Q_2 \triangleq q_{{}^2_0a}$ to $Q_3$ are equivalent to the stopping of subpopulation $\mc{A}_b$ or $\mc{A}_a$, respectively, at the obtained switching time $t_3$. Hence equation \eqref{eq:minorStoppCond} must also hold at the associated $t_3$ for each of the mentioned cases. Similarly, for the transition from $Q_2\triangleq q_{{}_0^1}$ to $Q_3$ both \eqref{majorHcontinuity_hybrid} and \eqref{MinorHcontinuity} must hold at the same time. 

We observe that if  \eqref{majorHcontinuity_hybrid} does not hold for any of the realizations of $Q_2=\{q_{{}^2_0a}, q_{{}_0^1}, q_{{}^2_0b}   \}$, then we conclude that $Q_3$ is not the final discrete state of the system. Subsequently, we start from \textit{Step 2} solving the dynamic programming backward in time from $t=T$. 

\textit{Step 2}. \textit{(Solving backwards for transitions from $Q_2$ to $Q_1$)}. Starting from the obtained realizations of $Q_2$ in \textit{Step 1} and the corresponding switching times $t_3$, we follow a similar approach as in \textit{Step1} to determine the realizations of $Q_1$ which may take place and their corresponding switching times $t_2$.  More specifically, equation \eqref{RiccatiMajor} is solved with the boundary (terminal) condition \eqref{MajorBoundaryC} with $j=3$ at $t_3$. Then, for example,  to determine from $Q_2 \triangleq q_{{}_0^2b}$ which of (either of or neither of) the transitions to $Q_1 \triangleq q_{{}_0^2ab}$  and $Q_1 \triangleq q_{{}_0^1b}$ may occur, equations \eqref{majorHcontinuity_hybrid}, \eqref{eq:minorStoppCond} and \eqref{majorHcontinuity_hybrid}, \eqref{MinorHcontinuity} are checked, respectively. 

\textit{Step 3}. \textit{(Solving backwards for transitions from $Q_1$ to $Q_0$)}. %\ali{I have modified Steps 3 (changing $t_1$ to $t_2$ and Step 4 (adding the evaluation of $t_1$). Please confirm that you agree}\dena{confirmed}
Similar to the previous steps, starting from the cases for $Q_1$ and the value for $t_2$ determined in \textit{Step 2}, it is investigated using equations \eqref{majorHcontinuity_hybrid}, \eqref{eq:minorStoppCond} and \eqref{MinorHcontinuity} whether the transition to $Q_0$ may occur or not and, if affirmative, the corresponding switching time $t_1$ is calculated.

\textit{Step 4}. \textit{(Specifying the optimal discrete sequence)}. In case the \textit{Steps 1-3} yield more than one sequence for the discrete state trajectory, the optimal sequence is determined by comparing the value functions along the obtained discrete state sequences with the value function for the case where no switching or stopping event happens. Finally it should be noted that if \textit{Steps 1-3} result in no realized discrete trajectory, then the system may remain in the discrete state $Q_0$ over the interval $[0,T]$.
\section{Simulations}
\label{sec:SimulationHybridMFG}
The framework introduced in the current paper can be applied to practical examples. In particular, an application to optimal execution problems in financial markets is indicated in reference \cite{FirooziPakniyatCainesCDC2017}. However, the case study in this section has been chosen to clearly illustrate the dynamical properties of hybrid MFG systems, in particular, how the stopping of one or both subpopulations may affect other agents in the game. Consider a system of 100 minor agents with two types $\mathcal{A}^a$ and $\mathcal{A}^b$ and a single major agent $\mathcal{A}_0$. {\color{black}The minor agents are provided with the option to quit if it is optimal for them to do so. Since the major agent is not permitted to switch in this example, the system has three discrete states $Q_0, Q_1, Q_2$, which index the stopping of one or both subpopulations as illustrated in Figure \ref{fig:HybridAutomataStop}.}
\begin{figure}
\centering
\scalebox{.45}{
\begin{tikzpicture}[node distance=27mm, auto]

  %%% first column %%%
  \node[state,minimum size=70pt](s123)
{$\begin{array}{c}
\mfs{$\bm{q_{{}_0^1ab}}$} \\
\\
\mfs{${\cal A}_{0}^{1}$},\\
\mfs{${\cal A}_{a},{\cal A}_{b}$}
\end{array}$};
 
  %%% second column %%%
 \node[state,minimum size=70pt,yshift=-2cm](s13) [right=20mm of s123]
%[left of=s23]
{$\begin{array}{c}
\mfs{$\bm{q_{{}_0^1a}}$} \\
\\
\mfs{${\cal A}_{0}^{1}$},\\
\mfs{${\cal A}_{a}$}
\end{array}$};

  \node[state,minimum size=70pt,yshift=2cm](s12) [right=20mm of s123]
{$\begin{array}{c}
\mfs{$\bm{q_{{}_0^1b}}$} \\
\\
\mfs{${\cal A}_{0}^{1}$},\\
\mfs{${\cal A}_{b}$}
\end{array}$};

  %%% third row %%%
  \node[state,minimum size=70pt] (s1)[right=67mm of s123] %[below=25mm of s23]
{$\begin{array}{c}
\mfs{$\bm{q_{{}_0^1}}$} \\
\\
\mfs{${\cal A}_{0}^{1}$}\\{ }
\end{array}$};

\path[-{Latex[scale=1.2]}]
(s123) edge[bend left=25, double, pos=0.7]  node {} (s12)
%(s123) edge[bend left=20]  node {$\sigma_{s_0}$} (s23)
(s123) edge[bend right=25, below, double]  node {} (s13)

%(s23) edge[bend right=15, double]  node {$\sigma_{s_a}$} (s2)
%(s23) edge[bend left=15, pos=0.32, double]  node {$\sigma_{s_b}$} (s3)

%(s12) edge[bend left=20]  node {$\sigma_{s_0}$} (s2)
(s12) edge[bend left=25, pos=0.3, double]  node {} (s1)

%(s13) edge[bend right=20]  node {$\sigma_{s_0}$} (s3)
(s13) edge[bend right=25,below, pos=0.45, double]  node {} (s1)

%(s1) edge[bend left=20]  node {$\sigma_{s_0}$} (s0)
%(s2) edge[bend left, double, pos=0.35]  node {$\sigma_{s_b}$} (s0)
%(s3) edge[bend right, double]  node {$\sigma_{s_a}$} (s0)

;

%\coordinate (q0) at ($(s123) + (0,0)$);
\node (A) [draw=magenta, fit= (s123), inner sep=3mm, 
            dashed, ultra thick, fill=magenta!10, fill opacity=0] {};
\node [xshift=3ex, yshift=9ex, magenta] at (A.west) {\mfs{$Q_0$}};
%\node [xshift=-5ex, yshift=2ex, white] at (A.west) {$[t_0,t_1)$};

\node (B) [draw=myGrn, fit= (s12) (s13), inner sep=3mm, 
            dashed, ultra thick, fill=myGrn!10, fill opacity=0] {};
\node [xshift=3ex, yshift=21ex, myGrn] at (B.west) {\mfs{$Q_1$}};
%\node [xshift=-5ex, yshift=2ex, white] at (B.west) {$[t_1,t_2)$};
    
\node (C) [draw=blue, fit= (s1) , inner sep=3mm, 
            dashed, ultra thick, fill=blue!10, fill opacity=0] {};
\node [xshift=-3ex, yshift=8ex, blue] at (C.east) {\mfs{$Q_2$}};
%\node [xshift=-5ex, yshift=2ex, white] at (C.west) {$[t_2,t_3)$};

%\node (D) [draw=orange, fit= (s0), inner sep=3mm, 
%            dashed, ultra thick, fill=orange!10, fill opacity=0] {};
%\node [xshift=3ex, yshift=7ex, orange] at (D.west) {$Q_3$};
%\node [xshift=-5ex, yshift=2ex, white] at (D.west) {$[t_3,T]$};

\end{tikzpicture}
}
\caption{{\color{black}Hybrid Automata Diagram with a single major agent and two populations of minor agents with stopping times.}} \label{fig:HybridAutomataStop}
\end{figure}

To clearly depict the impact of subpopulions' stopping on the control action and the trajectory of other agents, time-varying system matrices for a generic minor agent in subpopulation $\mathcal{A}^a$, with $N_a = 50$, are defined as
\[ {A}_a \triangleq \left[ \begin{array}{cc}
       2e^{-t} & e^{-0.5t} \\
        e^{-0.5t} & 2 e^{-t}  \end{array} \right],\quad  B_a \triangleq \left[ \begin{array}{cc}
        1 \\
        0.1 \end{array} \right], \]
and for a generic minor agent in subpopulation $\mathcal{A}^b$, with $N_b = 50$, are given by
\[ {A}_b \triangleq \left[ \begin{array}{cc}
       5 e^{-1.5t} \cos(t)  & 5 e^{-2t} \\
        5 e^{-2t} \sin(t)  & 5 e^{-1.5 t}  \end{array} \right],\quad  B_b \triangleq \left[ \begin{array}{cc}
        0 \\
        0.1 \end{array} \right], \]
and for the major agent are given by
\[ A_0 \triangleq \left[ \begin{array}{cc}
       2 e^{-t} & e^{-t} \\
       e^{-0.5t} & 2e^{-0.5t}  \end{array} \right], \quad B_0 \triangleq \left[ \begin{array}{cc}
       0.1 \\
        0.1 \end{array} \right]. \]
The parameters used in the simulation are: $ t_{final} = 18 \,\text{sec}$, $\Delta t = 0.01\, \text{sec},\,\sigma_0 = 0.015,\, \sigma_a = \sigma_b = 0.05$, $H_0 = 0.6\times I_{2}$, $H_1^a = H_1^b = 0.2\times I_{2},\, H_2^a = H_2^b = 0.02 \times I_{2}$, $G_a = G_b = 0_{2\times 2}$. We note that Assumptions \ref{ass:major_parameters}-\ref{ass:convexity_minor} are satisfied as the parameters are bounded analytic functions of time. Then following the steps in Section \ref{sec:HDP} the optimal control actions and state trajectories for a single realization in discrete states $Q_0$, $Q_1$, $Q_2$ for the entire population of $101$ agents are obtained (To solve the corresponding Riccati equations we use the MATLAB code\footnote{https://www.mathworks.com/matlabcentral/answers/94722-how-can-i-solve-the-matrix-riccati-differential-equation-within-matlab} recommended by MathWorks.). In Figure \ref{fig:hybridCntrls} and Figure \ref{fig:hybridTrajs} only 10 minor agents are shown for the sake of clarity. The subpopulations $\mc{A}^b$ and $\mc{A}^a$ stop, respectively, at $t_1=6$ sec and $t_2=12$ sec. This, in particular, impacts the behaviour of the major agent at the stopping times $t_1$ and $t_2$ as illustrated in Figure \ref{fig:hybridCntrls} and Figure \ref{fig:hybridTrajs}. 
\begin{figure}
%\vspace{-1.1cm}
\begin{center}
\includegraphics[width=0.8\linewidth]{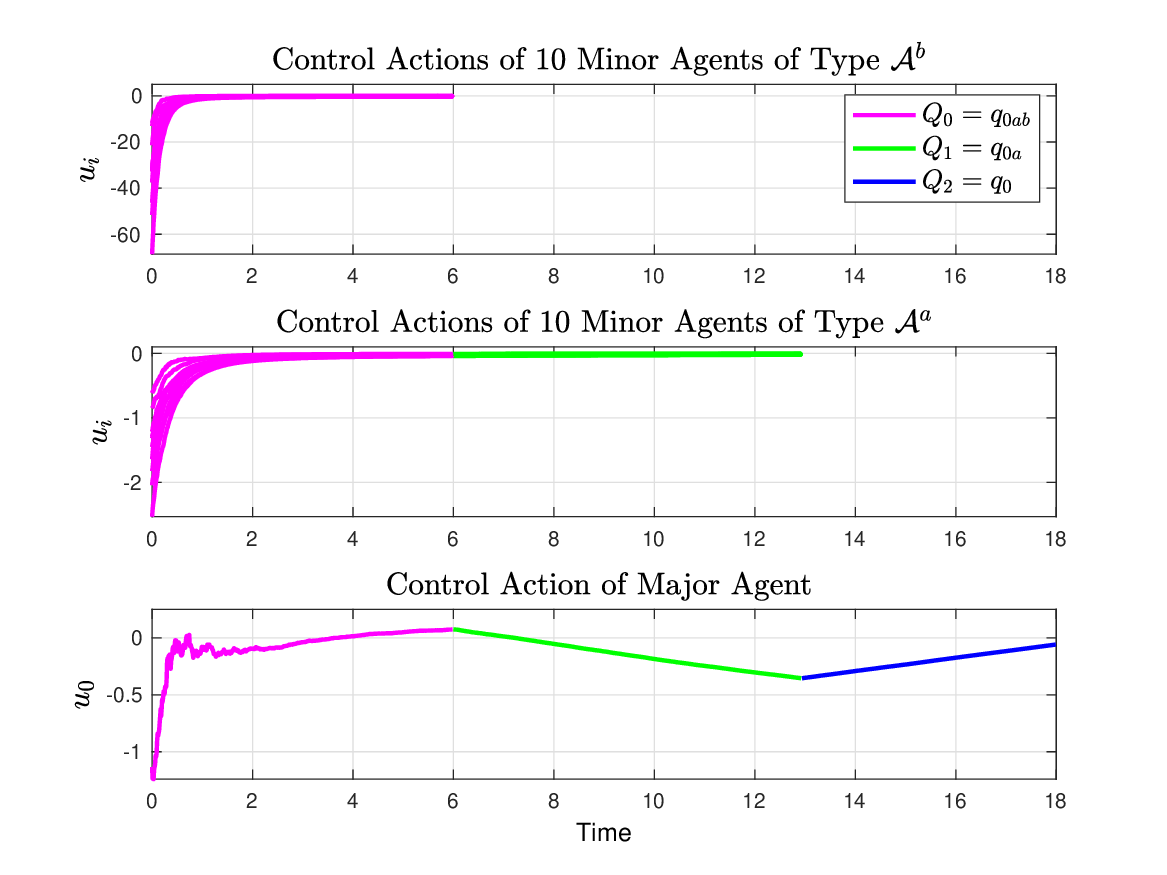}
\end{center}
%\vspace{-0.5cm}
\caption{The control actions for a single realization of the major agent, 10 sample minor agents of type $\mathcal{A}^a$, and 10 sample minor agents of type $\mathcal{A}^b$ in discrete states $Q_0$, $Q_1$, $Q_2$.}
\label{fig:hybridCntrls}
\end{figure}
\begin{figure}
%\vspace{-1cm}
\begin{center}
\includegraphics[width=0.8\linewidth]{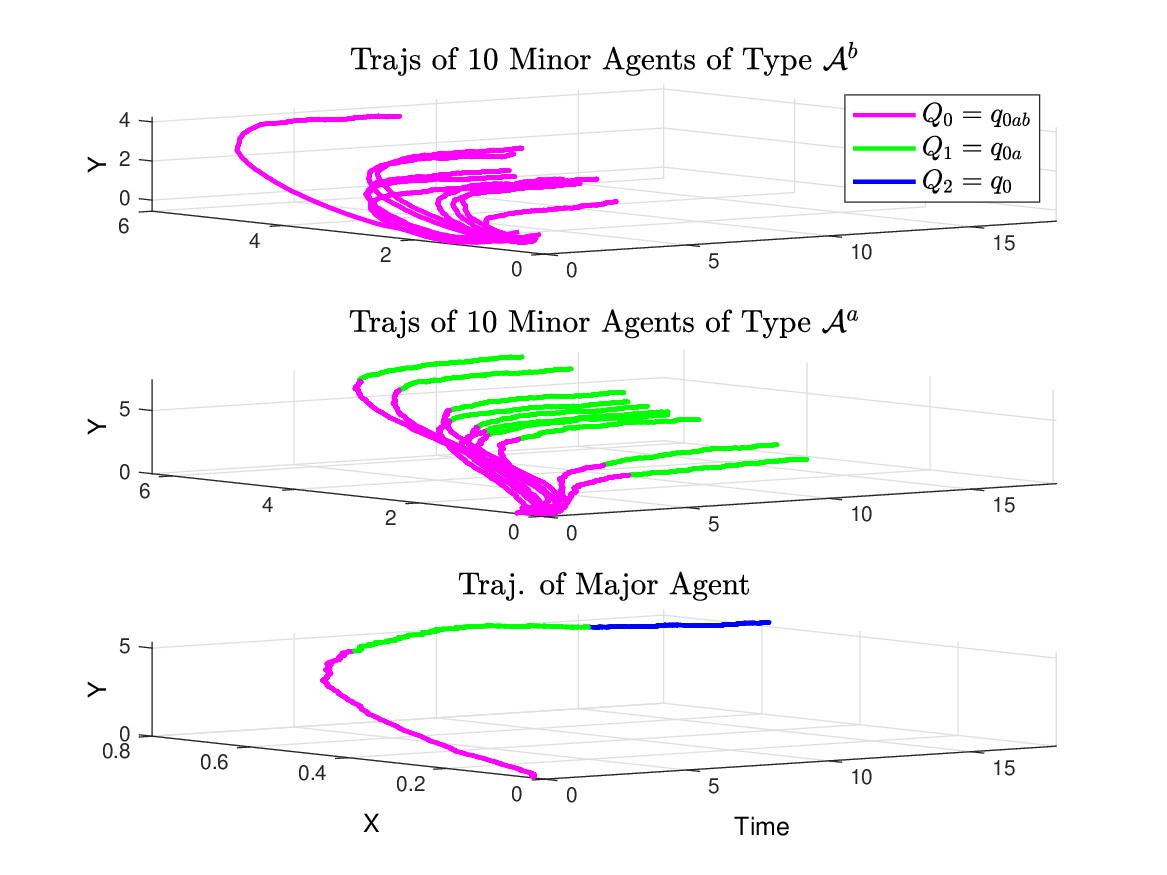}
\end{center}
%\vspace{-0.5cm}
\caption{The state trajectories for a single realization of the major agent, 10 sample minor agents of type $\mathcal{A}^a$, and 10 sample minor agents of type $\mathcal{A}^b$ in discrete states $Q_0$, $Q_1$, $Q_2$.} \label{fig:hybridTrajs}
\end{figure}

\section{Conclusions} \label{sec:ConclusionHybridMFG}
A hybrid mean field game theory has been established here for a class of major-minor LQG MFG systems for which controlled switching and stopping times are state and trajectory independent, and only depend on the dynamical and cost functional parameters of each agent. As a result, all agents of the same type would stop or switch at the same time. It is of significant interest to develop and extend the theory to account for switchings and stoppings at individual rates and/or upon arrival on switching manifolds, where individuals in subpopulations may quit or switch to alternative dynamics at different times. This is of particular importance in the modelling of optimal execution problems where traders stop or switch after reaching a specific number of shares. Another future direction is the tractable formulation for several subpopulations, including a systematic methodology for treating more complex discrete state sequence lattices.

%\appendix

%\ali{It shall be remarked that a general form of optimal stopping problems does not possess a Riccati expression of optimal inputs and, consequently, nonlinear versions of hybrid mean field games shall be formulated for such problems. However, as discussed in Theorem \ref{thm:HybridLQG} (and Corollary \ref{thm:HybridLQGstop}), for certain classes of hybrid and stopping LQG systems, the optimal switching (and stopping) times become state-invariant and, thus, can be deterministically identified.}

%\begin{ack}                               % Place acknowledgements
%Partially supported by the Roman Senate.  % here.
%\end{ack}

\bibliographystyle{plain}        % Include this if you use bibtex 
\bibliography{bib_Dena_08Oct20}           % and a bib file to produce the 
                                 % bibliography (preferred). The
                                 % correct style is generated by
                                 % Elsevier at the time of printing.

%\begin{thebibliography}{99}     % Otherwise use the  
                                 % thebibliography environment.
                                 % Insert the full references here.
                                 % See a recent issue of Automatica 
                                 % for the style.
%  \bibitem[Heritage, 1992]{Heritage:92}
%     (1992) {\it The American Heritage. 
%     Dictionary of the American Language.}
%     Houghton Mifflin Company.
%  \bibitem[Able, 1956]{Abl:56}
%     B.~C.~Able (1956). Nucleic acid content of macroscope. 
%     {\it Nature 2}, 7--9. 
%  \bibitem[Able {\em et al.}, 1954]{AbTaRu:54}   
%     B.~C. Able, R.~A. Tagg, and M.~Rush (1954).
%     Enzyme-catalyzed cellular transanimations.
%     In A.~F.~Round, editor, 
%     {\it Advances in Enzymology Vol. 2} (125--247). 
%     New York, Academic Press.
%  \bibitem[R.~Keohane, 1958]{Keo:58}
%     R.~Keohane (1958).
%     {\it Power and Interdependence: 
%     World Politics in Transition.}
%     Boston, Little, Brown \& Co.
%  \bibitem[Powers, 1985]{Pow:85}
%     T.~Powers (1985).
%     Is there a way out?
%     {\it Harpers, June 1985}, 35--47.

%\end{thebibliography}
\appendix
\section{Proof of Theorem \ref{thm:HybridLQG}}\label{proofThm1}
We invoke the Stochastic Hybrid Minimum Principle~\cite{APPECCDC2016} to form the family of system Hamiltonians as
\begin{multline}
H^{Q_j}\left(x^{Q_j},u^{Q_j},\lambda^{Q_j},K^{Q_j}\right)=\frac{1}{2}\Big(\left\Vert x^{Q_j}\left(t\right)\right\Vert_{P^{Q_j}\left(t\right)}^{2}\\+\left\Vert u^{Q_j}\left(t\right)\right\Vert_{R^{Q_j}\left(t\right)}^{2}\Big)
+\left[\lambda^{Q_j}\right]^{T}\!\left(A^{Q_j} x^{Q_j}+B^{Q_j} u^{Q_j}\right)\\\!+\left[K^{Q_j}\right]^{T} D^{Q_j},
\label{HamiltonianForm}
\end{multline}
It immediately follows that
\begin{equation}
\underset{u^{Q_j} \in \mathbb{R}^{m_{Q_j}}}{\text{argmin }} H^{Q_j}\left(x^{Q_j},u^{Q_j},\lambda^{Q_j},K^{Q_j}\right) = -\left[R^{Q_j}\right]^{-1} \left[B^{Q_j}\right]^T \lambda^{Q_j}.
\label{OptimalInputProof}
\end{equation}
%and, therefore, it remains to be shown that along a trajectory $x^{Q_j}\left(t\right)$ associated with the input \eqref{OptimalFeedbackLaw_switch} and switchings at $t_j$'s satisfying \eqref{DeltaHzero}--\eqref{DeltaHnegative}, the
In order to find conditions under which the adjoint processes take the form $\lambda^{Q_j}\left(t\right) = \Pi^{Q_j}\left(t\right) x^{Q_j}\left(t\right)$ %are adjoint processes of the associated optimal control problem.
%
%Beginning 
we begin 
with the final discrete state $Q_L$, and follow similar arguments as those in classical LQG theory (see e.g. \cite[Chapter 6]{XYZBook1999}), %it is shown that $\lambda^{Q_{L}}\left(t\right) = \Pi^{Q_{L}}\left(t\right) x^{Q_{L}}\left(t\right)$ as
to obtain
\begin{align}
\lambda^{Q_L}\left(T\right) &= \Pi^{Q_L}\left(t\right) x^{Q_L}\left(T\right) = \frac{1}{2}\frac{\partial}{\partial x} \left\Vert x\left(T\right)\right\Vert_{\bar{P}^{Q_{L}}(T)}^{2},\\
d\lambda^{Q_L}=&-\frac{\partial H^{Q_L}}{\partial x}\left(x^{Q_L},u^{Q_L},\lambda^{Q_L},K^{Q_L}\right)dt+K^{Q_L} dw \nonumber\\
&=-\left(P^{Q} x^{Q_L}+\left[A^{Q_L}\right]^{T}\lambda^{Q_L}\right)dt+K^{Q_L}dw \label{dlambdaLQG},
\end{align}
with $K^{Q_L}\left(t\right) = \Pi^{Q_L}\left(t\right) D^{Q_L}$.

Within the (backward) induction procedure, we hypothesize that $\lambda^{Q_{j+1}}\left(t\right) = \Pi^{Q_{j+1}}\left(t\right) x^{Q_{j+1}}\left(t\right)$ holds and determine conditions under which $\lambda^{Q_{j}}\left(t\right) = \Pi^{Q_{j}}\left(t\right) x^{Q_{j}}\left(t\right)$ is concluded. To this end, we note that from \cite[Theorem 1]{APPECCDC2016} (see also \cite{APPEC2017IFAC}) the adjoint processes and Hamiltonians must satisfy
\begin{align}
\lambda^{Q_{j}}\left(t_{j+1}\right) =\left[\Psi_{\sigma_{j}}\right]^{T} \lambda^{Q_{j+1}}\left(t_{j+1}+\right) &+ C_{\sigma_{j}}x^{Q_j}(t_{j+1}) , \label{LQStochasticAdjointBoundaryCondition}\allowdisplaybreaks\\
H^{Q_{j}}_{^{\big(x^{Q_{j}},u^{o,Q_{j}},\lambda^{Q_{j}},K^{Q_{j}}\big)}} - \left[K^{Q_{j}}\right]^T D^{Q_{j}}  -& \frac{1}{2} \frac{\partial}{\partial t}\left\Vert x^{Q_{j}}\right\Vert^{2}_{C^{\left(t\right)}_{\sigma_{j}}}\bigg|_{t^\omega_{j+1}-}
\nonumber\allowdisplaybreaks\\
= H^{Q_{j+1}}_{^{\big(x^{Q_{j+1}},u^{o,Q_{j+1}},\lambda^{Q_{j+1}},K^{Q_{j+1}}\big)}} - &\left[K^{Q_{j+1}}\right]^{T} D^{Q_{j+1}} \bigg|_{t^\omega_{j+1}}.
\label{HamiltonianJump}
\end{align}
The substitution of the hypothesized Riccati forms into \eqref{LQStochasticAdjointBoundaryCondition} yields
\begin{multline}
    \Pi^{Q_{j}}\left(t_{j+1}\right)x^{Q_{j}}\left(t_{j+1}-\right)
    \\
    =\left[\Psi_{\sigma_{j}}\right]^{T}\Pi^{Q_{j+1}}\left(t_{j+1}\right)x^{Q_{j+1}}\left(t_{j+1}\right)+C_{\sigma_{j}}x^{Q_{j}}(t_{j+1}-)
    \\
    =\left[\Psi_{\sigma_{j}}\right]^{T}\Pi^{Q_{j+1}}\left(t_{j+1}\right)\Psi_{\sigma_{j}}x^{Q_{j}}(t_{j+1}-)+C_{\sigma_{j}}x^{Q_{j}}(t_{j+1}-)
    \label{PreStochasticRiccatiBC}
\end{multline}
and, therefore, if \eqref{StochasticRiccatiBC} holds, then \eqref{PreStochasticRiccatiBC} is obtained regardless of the value of $x^{Q_{j}}(t_{j+1}-)$. Moreover, the substitution of $\lambda^{Q}\left(t\right) = \Pi^{Q}\left(t\right) x^{Q}\left(t\right)$ into \eqref{OptimalInputProof}, and subsequently in \eqref{HamiltonianForm}, yields
%\begin{multline}
%    H^{Q}\left(x,u^{o},\lambda,K^{Q}\right)-\left[K^{Q}\right]^{T}D^{Q}
%    \allowdisplaybreaks\\
%    =\frac{1}{2}\left\Vert x\right\Vert _{P^{Q}+A^{Q}\Pi^{Q}+\Pi^{Q}\left[A^{Q}\right]^{T}-\Pi^{Q}B^{Q}\left[R^{Q}\right]^{-1}\left[B^{Q}\right]^{T}\Pi^{Q}}^{2} \label{HamiltonianOptimized}
%\end{multline}
%%% the version that will be used in the final paper
\begin{multline}
   H^{Q}\left(x,u^{o},\lambda,K^{Q}\right)-\left[K^{Q}\right]^{T}D^{Q}
    \allowdisplaybreaks\\
    =\frac{1}{2}\left\Vert x\right\Vert _{P^{Q}+\Pi^{Q}A^{Q}+\left[A^{Q}\right]^{T}\Pi^{Q}-\Pi^{Q}B^{Q}\left[R^{Q}\right]^{-1}\left[B^{Q}\right]^{T}\Pi^{Q}}^{2} \label{HamiltonianOptimized}
\end{multline}
and in particular, at the switching instant $t_{j+1}$, the substitution of \eqref{HamiltonianOptimized} and \eqref{JumpMap} into \eqref{HamiltonianJump} result in
%\begin{multline}
%\frac{1}{2}[x^{Q_{j}}(t_{j+1}-)]^{T}\bigg(P^{Q_{j}}+A^{Q_{j}}\Pi_{(t_{j+1})}^{Q_{j}}+\Pi_{(t_{j+1})}^{Q_{j}}\left[A^{Q_{j}}\right]^{T}
%\allowdisplaybreaks\\
%-\Pi_{(t_{j+1})}^{Q_{j}}B^{Q_{j}}\left[R^{Q_{j}}\right]^{-1}\left[B^{Q_{j}}\right]^{T}\Pi_{(t_{j+1})}^{Q_{j}}\bigg)x^{Q_{j}}(t_{j+1}-)
%\allowdisplaybreaks\\
%=\frac{1}{2}[x^{Q_{j+1}}(t_{j+1})]^{T}\bigg(P^{Q_{j+1}}+A^{Q_{j+1}}\Pi_{(t_{j+1})}^{Q_{j+1}}+\Pi_{(t_{j+1})}^{Q_{j+1}}\left[A^{Q_{j+1}}\right]^{T}
%\allowdisplaybreaks\\
%-\Pi_{(t_{j+1})}^{Q_{j+1}}B^{Q_{j+1}}\left[R^{Q_{j+1}}\right]^{-1}\left[B^{Q_{j+1}}\right]^{T}\Pi_{(t_{j+1})}^{Q_{j+1}}\bigg)x^{Q_{j+1}}(t_{j+1})
%\allowdisplaybreaks\\
%=\frac{1}{2}[\Psi_{\sigma_{j}}x^{Q_{j}}_{(t_{j+1}-)}]^{T}\bigg(P^{Q_{j+1}}+A^{Q_{j+1}}\Pi_{(t_{j+1})}^{Q_{j+1}}+\Pi_{(t_{j+1})}^{Q_{j+1}}\left[A^{Q_{j+1}}\right]^{T}
%\allowdisplaybreaks\\
%-\Pi_{(t_{j+1})}^{Q_{j+1}}B^{Q_{j+1}}\left[R^{Q_{j+1}}\right]^{-1}\left[B^{Q_{j+1}}\right]^{T}\Pi_{(t_{j+1})}^{Q_{j+1}}\bigg)\Psi_{\sigma_{j}}x^{Q_{j}}_{(t_{j+1}-)}
%\label{HamiltonianContinuityExpand}
%\end{multline}
%%% the version that will be used in the final paper
\begin{multline}
\frac{1}{2}[x^{Q_{j}}(t_{j+1}-)]^{T}\bigg(P^{Q_{j}}+\Pi_{(t_{j+1})}^{Q_{j}}A^{Q_{j}}+\left[A^{Q_{j}}\right]^{T}\Pi_{(t_{j+1})}^{Q_{j}}
\allowdisplaybreaks\\
-\Pi_{(t_{j+1})}^{Q_{j}}B^{Q_{j}}\left[R^{Q_{j}}\right]^{-1}\left[B^{Q_{j}}\right]^{T}\Pi_{(t_{j+1})}^{Q_{j}}\\-\frac{\partial{C}_{\sigma_j}(t)}{\partial t}{\bigg|}_{t=t_{j+1}}\bigg)x^{Q_{j}}(t_{j+1}-)
\allowdisplaybreaks\\
=\frac{1}{2}[x^{Q_{j+1}}(t_{j+1})]^{T}\bigg(P^{Q_{j+1}}+\Pi_{(t_{j+1})}^{Q_{j+1}}A^{Q_{j+1}}+\left[A^{Q_{j+1}}\right]^{T}\Pi_{(t_{j+1})}^{Q_{j+1}}
\allowdisplaybreaks\\
-\Pi_{(t_{j+1})}^{Q_{j+1}}B^{Q_{j+1}}\left[R^{Q_{j+1}}\right]^{-1}\left[B^{Q_{j+1}}\right]^{T}\Pi_{(t_{j+1})}^{Q_{j+1}}\bigg)x^{Q_{j+1}}(t_{j+1})
\allowdisplaybreaks\\
=\frac{1}{2}[\Psi_{\sigma_{j}}x^{Q_{j}}_{(t_{j+1}-)}]^{T}\bigg(P^{Q_{j+1}}+\Pi_{(t_{j+1})}^{Q_{j+1}}A^{Q_{j+1}}+\left[A^{Q_{j+1}}\right]^{T}\Pi_{(t_{j+1})}^{Q_{j+1}}
\allowdisplaybreaks\\
-\Pi_{(t_{j+1})}^{Q_{j+1}}B^{Q_{j+1}}\left[R^{Q_{j+1}}\right]^{-1}\left[B^{Q_{j+1}}\right]^{T}\Pi_{(t_{j+1})}^{Q_{j+1}}\bigg)\Psi_{\sigma_{j}}x^{Q_{j}}_{(t_{j+1}-)}. 
\label{HamiltonianContinuityExpand}
\end{multline}
In particular, if \eqref{DeltaHzero} holds, then \eqref{HamiltonianContinuityExpand} holds independent of the realization for $x^{Q_{j}}(t_{j+1}-)$ and thus, independent of ${\cal F}_t$.
Since the satisfaction of \eqref{StochasticRiccatiBC} and \eqref{DeltaHzero} lead to the satisfaction of \eqref{LQStochasticAdjointBoundaryCondition} and \eqref{HamiltonianJump} with ${\cal F}_t$-independence, then the induction hypothesis is true due to the uniqueness of the solution to the backward differential equations \eqref{dlambdaLQG}. Moreover, \eqref{DeltaHpositive} and \eqref{DeltaHnegative} ensure that such a switching instant is unique and therefore the associated Riccati equations and switching conditions globally correspond to a unique optimal strategy.
\section{Jump Transition Maps and Switching Costs for Minor Agents} \label{sec:minorJumpMapNSwitchingCost}
We define the notation $M_k^{Q_j}(l:m), \, k \triangleq a,b, \, 0\leq j \leq 3$, which shall be used to identify the switching cost associated with switching time $t_j,\, 1 \leq j \leq 3$, i.e., the cost incurred when a change in the system happens. Matrix $M_{k}^{Q_j}(l:m)$ is made by making all the entires of $\bar{\mathbb{P}}^{Q_j}_k$ zero except those associated with its $l$-th to $m$-th columns and rows, hence it has the same size as $\bar{\mathbb{P}}^{Q_j}_k$, i.e.    
%\vspace{0.03cm}
\begin{align}
%\overbrace{}^{\text{$l$ to $m$ th columns of} \, \, \bar{\mathbb{P}}_k^{Q_j} }
M_{k}^{Q_j}(l:m) = \left[ 
\begin{array}{c|c|c}
\bovermat{$\bar{\mathbb{P}}_k^{Q_j}(:,l:m)$}{0&\hspace{0.6cm}} &0 \\
\hline
%\vspace{0.01cm}
& &\\
\hline 
0&  &0
\end{array} \right]_{\text{size}(\scriptscriptstyle{\bar{\mathbb{P}}_k^{Q_j}})} \begin{aligned}
 % &\left.\begin{matrix}
%  \partialphantom m = 1  \\[0.5em]
 % \partialphantom m = 2  \\[0.5em]
%  \cdots \\[0.5em]
%  \partialphantom m = M  \\[0.5em]
%  \end{matrix} \right\} %
%  p = 1\\
  %&\begin{matrix}
 % \\[-1.67em]\phantom{\cdots}\cdots
 % \end{matrix}\\ %
 &\left.
 \begin{matrix}
  %\partialphantom m = 2  \\[0.5em]
 \hspace{-2cm} \\[0.75em] 
  %\cdots \\[0.5em]  
  \end{matrix} \hspace{-1cm}\right\}
 \scriptstyle{ \bar{\mathbb{P}}_k^{Q_j}(l\,:\,m,\,:)}\\ 
 \end{aligned} 
 \end{align}     
 where $\bar{\mathbb{P}}_k^{Q_j}(:,l:m)$ and $\bar{\mathbb{P}}_k^{Q_j}(l:m,:)$, respectively, denote $l$-th to $m$-th columns and rows of $\bar{\mathbb{P}}_k^{Q_j}$.
 
The values of minor agent $\mathcal{A}_i^k$ continuous state before and after the switching time $t_1$ satisfy the following jump transition map 
\begin{align}\label{JumpTranMinorMap3}
 x_{i}^{ex,Q_{1}}(t_{1})=\Psi_{i,\sigma_1}^k x_{i}^{ex,Q_{0}}(t_{1}-),
 \end{align}
where for $k \triangleq a$
\begin{align}
\Psi_{i,\sigma_1}^a = \begin{cases}
\Psi_{i,\sigma_{q_{{}_0^1ab},q_{{}_0^2ab}}}^a = I_{3n \times 3n}, \\%& \hspace{0.5cm} \text{for transition from $Q_0 =q_{{}_0^1ab}$ to $Q_1 = q_{{}_0^2ab}$,}\\
\Psi_{i,\sigma_{q_{{}_0^1ab},q_{{}_0^1a}}}^a = \left[ \begin{array}{ccc} 
 I_n & 0_{n \times n} & 0_{n \times n}\\
 0_{n \times n} & I_n & 0_{n \times n} \end{array} \right],\\%& \hspace{0.5cm} \text{for transition from $Q_0 =q_{{}_0^1ab}$ to $Q_1 = q_{{}_0^1a}$,}\\
\Psi_{i,\sigma_{q_{{}_0^1ab},q_{{}_0^1b}}}^a = \left[ \begin{array}{ccc} 
 0_{n \times n} & 0_{n \times n} & 0_{n \times n}
 \end{array} \right].% & \hspace{0.5cm} \text{for transition from $Q_0 =q_{{}_0^1ab}$ to $Q_1 = q_{{}_0^1b}$.}
\end{cases}
\end{align}
% \begin{align}
% \Psi_{i,\sigma_1}^a = \begin{cases}
%  I_{3n \times 3n}, \text{for $\sigma_1 = \sigma_{q_{{}_0^1ab},q_{{}_0^2ab}}$}\\%& \hspace{0.5cm} \text{for transition from $Q_0 =q_{{}_0^1ab}$ to $Q_1 = q_{{}_0^2ab}$,}\\
% \Psi_{i,\sigma_{q_{{}_0^1ab},q_{{}_0^1a}}}^a = \left[ \begin{array}{ccc} 
%  I_n & 0_{n \times n} & 0_{n \times n}\\
%  0_{n \times n} & I_n & 0_{n \times n} \end{array} \right],\\%& \hspace{0.5cm} \text{for transition from $Q_0 =q_{{}_0^1ab}$ to $Q_1 = q_{{}_0^1a}$,}\\
% \Psi_{i,\sigma_{q_{{}_0^1ab},q_{{}_0^1b}}}^a = \left[ \begin{array}{ccc} 
%  0_{n \times n} & 0_{n \times n} & 0_{n \times n}
%  \end{array} \right].% & \hspace{0.5cm} \text{for transition from $Q_0 =q_{{}_0^1ab}$ to $Q_1 = q_{{}_0^1b}$.}
% \end{cases}
% \end{align}
Moreover, the weight matrix $\mathbb{C}^a_{i,\sigma_1}$ associated with the switching cost in \eqref{minorCostQuadraticHyb} at time $t_1$ is specified as     
\begin{align}
\mathbb{C}_{i,\sigma_1}^a = \begin{cases}
\mathbb{C}_{i,\sigma_{q_{{}_0^1ab},q_{{}_0^2ab}}}^a = 0_{3n \times 3n},\\% & \hspace{0.5cm} \text{for transition from $Q_0 =q_{{}_0^1ab}$ to $Q_1 = q_{{}_0^2ab}$,}\\
\mathbb{C}_{i,\sigma_{q_{{}_0^1ab},q_{{}_0^1a}}}^a = M_a^{q_{{}_0^1ab}}(3n+1:4n),\\ %& \hspace{0.5cm} \text{for transition from $Q_0 =q_{{}_0^1ab}$ to $Q_1 = q_{{}_0^1a}$,}\\
\mathbb{C}_{i,\sigma_{q_{{}_0^1ab},q_{{}_0^1b}}}^a = \bar{\mathbb{P}}^{q_{{}_0^1ab}}_a. %& \hspace{0.5cm} \text{for transition from $Q_0  =q_{{}_0^1ab}$ to $Q_1 = q_{{}_0^1b}$.}
\end{cases}
\end{align}
For $k \triangleq b$, the jump transition map \eqref{JumpTranMinorMap3} at $t_1$ is given by
\begin{align}
\Psi_{i,\sigma_1}^b = \begin{cases}
\Psi_{i,\sigma_{q_{{}_0^1ab}q_{{}_0^2ab}}}^b = I_{3n \times 3n},\\%& \hspace{0.5cm} \text{for transition from $Q_0 =q_{{}_0^1ab}$ to $Q_1 = q_{{}_0^2ab}$,}\\
\Psi_{i,\sigma_{q_{{}_0^1ab},q_{{}_0^1a}}}^b = \left[ \begin{array}{ccc} 
 0_{n \times n} & 0_{n \times n} & 0_{n \times n}
 \end{array} \right],\\% & \hspace{0.5cm} \text{for transition from $Q_0 =q_{{}_0^1ab}$ to $Q_1 = q_{{}_0^1a}$,}\\
\Psi_{i,\sigma_{q_{{}_0^1ab},q_{{}_0^1b}}}^b = \left[ \begin{array}{ccc} 
 I_n & 0_{n \times n} & 0_{n \times n}\\
 0_{n \times n} & 0_{n \times n} & I_n \end{array} \right],\\% & \hspace{0.5cm} \text{for transition from $Q_0  =q_{{}_0^1ab}$ to $Q_1 = q_{{}_0^1b}$,}
 \end{cases}
\end{align}
and the corresponding switching cost weight matrix is 
\begin{equation}%\abovedisplayskip\abovedisplayshortskip
\mathbb{C}_{i,\sigma_1}^b = \begin{cases}
\mathbb{C}_{i,\sigma_{q_{{}_0^1ab}q_{{}_0^2ab}}}^b = I_{3n \times 3n},\\% & \hspace{0.5cm} \text{for transition from $Q_0 =q_{{}_0^1ab}$ to $Q_1 = q_{{}_0^2ab}$,}\\
\mathbb{C}_{i,\sigma_{q_{{}_0^1ab}q_{{}_0^1a}}}^b = \bar{\mathbb{P}}_b^{q_{{}_0^1ab}},\\%& \hspace{0.5cm} \text{for transition from $Q_0 =q_{{}_0^1ab}$ to $Q_1 = q_{{}_0^1a}$,}\\
\mathbb{C}_{i,\sigma_{q_{{}_0^1ab}q_{{}_0^1b}}}^b = M_b^{q_{{}_0^1ab}q_{{}_0^1b}}(2n+1:3n).% & \hspace{0.5cm} \text{for transition from $Q_0  =q_{{}_0^1ab}$ to $Q_1 = q_{{}_0^1b}$.}
\end{cases}
\end{equation}
The values of the minor agent's continuous state before and after the switching at $t_2$ satisfy the following jump map 
\begin{align}\label{JumpTranMinorMap2}
 x_{i}^{ex,Q_{2}}(t_{2})=\Psi_{i,\sigma_2}^k x_{i}^{ex,Q_{1}}(t_{2}-),
 \end{align}
where $\Psi_{i,\sigma_2}^k,\,k \triangleq a$, is given by
\begin{align}
\Psi_{i,\sigma_2}^a = \begin{cases}
\Psi_{i,\sigma_{q_{{}_0^1a}q_{{}_0^2a}}}^a= I_{2n \times 2n},\\%  &\hspace{0.5cm} \text{for transition from $Q_1 =q_{{}_0^1a} $ to $Q_2 = q_{{}_0^2a}$,} \\
\Psi_{i,\sigma_{q_{{}_0^1a}q_{{}_0^1}}}^a= \left[ \begin{array}{cc} 0_{n \times n} & 0_{n \times n}\end{array} \right],\\%&\hspace{0.5cm}\text{for transition from $Q_1 =q_{{}_0^1a} $ to $Q_2 = q_{{}_0^1}$,}\\
\Psi_{i,\sigma_{q_{{}_0^2ab}q_{{}_0^2a}}}^a= \left[ \begin{array}{ccc} 
I_n & 0_{n \times n} & 0_{n \times n} \\
0_{n \times n} & I_n & 0_{n \times n}
\end{array} \right],\\%&\hspace{0.5cm}\text{for transition from $Q_1 =q_{{}_0^2ab} $ to $Q_2 = q_{{}_0^2a}$,}\vspace{0.15cm}\\
\Psi_{i,\sigma_{q_{{}_0^2ab}q_{{}_0^2b}}}^a= \left[ \begin{array}{ccc}
 0_{n \times n} & 0_{n \times n} & 0_{n \times n}\\
 %0_{n \times n} & 0_{n \times n} & I_n 
 \end{array} \right].\\%&\hspace{0.5cm}\text{for transition from $Q_1 =q_{{}_0^2ab} $ to $Q_2 = q_{{}_0^2b}$.}
% \Psi_{i,q_{{}_0^1b}q_{{}_0^2b}}^a= I_{2n \times 2n},&\hspace{0.5cm}\text{for transition from $Q_1 =q_{{}_0^1b}$ to $Q_2 = q_{{}_0^2b}$,}\\
%\Psi_{i,q_{{}_0^1b}q_{{}_0^1}}^a= \left[ \begin{array}{cc}
%I_n & 0_{n \times n}
%\end{array} \right], &\hspace{0.5cm}\text{for transition from $Q_1 =q_{{}_0^1b}$ to $Q_2 = q_{{}_0^1}$,}
\end{cases}
\end{align}
Furthermore, the weight matrix $\mathbb{C}_{i,\sigma_2}^a$ associated with the switching cost at time $t_2$ is specified by 
\begin{align}
\mathbb{C}_{i,\sigma_2}^a = \begin{cases}
\mathbb{C}_{i,\sigma_{q_{{}_0^1a},q_{{}_0^2a}}}^a= 0_{2n \times 2n},\\%  &\hspace{0.5cm} \text{for transition from $Q_1 =q_{{}_0^1a} $ to $Q_2 = q_{{}_0^2a}$,} \\
\mathbb{C}_{i,\sigma_{q_{{}_0^1a},q_{{}_0^1}}}^a= \bar{\mathbb{P}}_a^{q_{{}_0^1a}},\\%&\hspace{0.5cm}\text{for transition from $Q_1 =q_{{}_0^1a} $ to $Q_2 = q_{{}_0^1}$,}\\
\mathbb{C}_{i,\sigma_{q_{{}_0^2ab},q_{{}_0^2a}}}^a= M_a^{q_{{}_0^2ab}}(3n+1:4n) ,\\%&\hspace{0.5cm}\text{for transition from $Q_1 =q_{{}_0^2ab} $ to $Q_2 = q_{{}_0^2a}$,}\vspace{0.15cm}\\
\mathbb{C}_{i,\sigma_{q_{{}_0^2ab},q_{{}_0^2b}}}^a= \bar{\mathbb{P}}_a^{q_{{}_0^2ab}}.
%&\hspace{0.5cm}\text{for transition from $Q_1 =q_{{}_0^2ab} $ to $Q_2 = q_{{}_0^2b}$.}
\end{cases}
\end{align}
In \eqref{JumpTranMap2}, the jump transition map $\Psi_{i,\sigma_2}^k,\, k \triangleq b$, %is given by
and the corresponding switching cost weight matrix $\mathbb{C}_{i,\sigma_2}^b $ %is
are given by
\begin{align}
\Psi_{i,\sigma_2}^b &= \begin{cases}
\Psi_{i,\sigma_{q_{{}_0^1b},q_{{}_0^2b}}}^b = I_{2n \times 2n},\\%  &\hspace{0.5cm} \text{for transition from $Q_1 =q_{{}_0^1a} $ to $Q_2 = q_{{}_0^2a}$,} \\
\Psi_{i,\sigma_{q_{{}_0^1b},q_{{}_0^1}}}^b = \left[ \begin{array}{cc} 0_{n \times n} & 0_{n \times n}\end{array} \right],\\%&\hspace{0.5cm}\text{for transition from $Q_1 =q_{{}_0^1a} $ to $Q_2 = q_{{}_0^1}$,}\\
\Psi_{i,\sigma_{q_{{}_0^2ab},q_{{}_0^2a}}}^b= \left[ \begin{array}{ccc}
 0_{n \times n} & 0_{n \times n} & 0_{n \times n}\\
 \end{array} \right],\\%&\hspace{0.5cm}\text{for transition from $Q_1 =q_{{}_0^2ab} $ to $Q_2 = q_{{}_0^2a}$,}\vspace{0.15cm}\\
\Psi_{i,\sigma_{q_{{}_0^2ab},q_{{}_0^2b}}}^b= \left[ \begin{array}{ccc} 
I_n & 0_{n \times n} & 0_{n \times n} \\
0_{n \times n} &  0_{n \times n} & I_n 
\end{array} \right].%&\hspace{0.5cm}\text{for transition from $Q_1 =q_{{}_0^2ab} $ to $Q_2 = q_{{}_0^2b}$,}
\end{cases}
\allowdisplaybreaks\\
\mathbb{C}_{i,\sigma_2}^b &= \begin{cases}
\mathbb{C}_{i,\sigma_{q_{{}_0^1b},q_{{}_0^2b}}}^b = 0_{2n \times 2n},\\%  &\hspace{0.5cm} \text{for transition from $Q_1 =q_{{}_0^1a} $ to $Q_2 = q_{{}_0^2a}$,} \\
\mathbb{C}_{i,\sigma_{q_{{}_0^1b},q_{{}_0^1}}}^b = \bar{\mathbb{P}}^{q_{{}_0^1b}}_b,\\%&\hspace{0.5cm}\text{for transition from $Q_1 =q_{{}_0^1a} $ to $Q_2 = q_{{}_0^1}$,}\\
\mathbb{C}_{i,\sigma_{q_{{}_0^2ab},q_{{}_0^2a}}}^b= \bar{\mathbb{P}}_b^{q_{{}_0^2ab}},\\% &\hspace{0.5cm}\text{for transition from $Q_1 =q_{{}_0^2ab} $ to $Q_2 = q_{{}_0^2a}$,}\vspace{0.15cm}\\
\mathbb{C}_{i,\sigma_{q_{{}_0^2ab},q_{{}_0^2b}}}^b= M_b^{q_{{}_0^2ab}}(2n+1:3n). %&\hspace{0.5cm}\text{for transition from $Q_1 =q_{{}_0^2ab} $ to $Q_2 = q_{{}_0^2b}$.}
\end{cases}
\end{align}
The values of the minor agent's continuous state before and after the switching at $t_3$ satisfy the jump transition map 
\begin{align}\label{JumpTranMapMinor3}
 x_{i}^{ex,Q_{3}}(t_{3})=\Psi_{i,\sigma_3}^k x_{i}^{ex,Q_{2}}(t_{3}-),
 \end{align}
where for $k \triangleq a$ 
\begin{gather}
\Psi_{i,\sigma_3}^a = \Psi^a_{i,\sigma_{q_{{}_0^2a},q_{{}_0^2}}} = \left[ \begin{array}{ccc}
0_{n \times n} & 0_{n \times n} &  0_{n \times n} \end{array} \right],\allowdisplaybreaks\\
\mathbb{C}_{i,\sigma_3}^a = \mathbb{C}^a_{i,\sigma_{q_{{}_0^2a}q_{{}_0^2}}} = \bar{\mathbb{P}}_a^{q_{{}_0^2a}},
\end{gather} 
and for $k \triangleq b$
\begin{gather}
\Psi^b_{i,\sigma_3} = \Psi^b_{i, \sigma_{q_{{}_0^2b}q_{{}_0^2}}} = \left[ \begin{array}{ccc} 
 0_{n \times n} & 0_{n \times n} &  0_{n \times n}\end{array} \right],\\
\mathbb{C}^b_{i,\sigma_3} = \mathbb{C}^b_{i, \sigma_{q_{{}_0^2b}q_{{}_0^2}}} = \bar{\mathbb{P}}_b^{q_{{}_0^2b}}.
\end{gather}

\section{Proof of Theorem \ref{thm:ENE_hybrid}}\label{app:NashProof}
The $\epsilon$-Nash property can be shown in two steps for the major agent and a generic minor agent, respectively.  Due to space limitation we detail the proof for the major agent and that of a minor agent will follow in the same manner.  
\begin{itemize}
    \item[(i)] Suppose that there exists a sequence $\{\delta_n \}_{n=1}^N$ such that $\delta_N \rightarrow 0$ as $N \rightarrow \infty$, and $\left| \tfrac{N_k}{N} -  \pi_k\right| = o(\delta_N)$, for all $k \triangleq a, b$. Given that all minor agents $\mc{A}_i^k,\, 1 \leq i \leq N,$ are using the optimal control actions $\mathcal{U}_{MF}^{N_t} \triangleq \{ {u}_i^{Q_j}; 1\leq i \leq N_t \},\, 1\leq N_t \leq N < \infty,$ given by \eqref{minorControl}, and the major agent is using an arbitrary control action $u_0 \in {\mc{U}}_{g}^{N,L}$, we show that  
    \begin{align}
      &(a)~~ \mb{E} \Vert x_0^{Q_j,N} - x_0^{Q_j} \Vert^2 \leq C( o(\tfrac{1}{N}) + o (\delta_N^2)),\label{conv_order_state}\allowdisplaybreaks\\
      &(b)~~ \mb{E} \Vert x^{(N^{Q_j})} - \bar{x}^{Q_j} \Vert^2 \leq C( o(\tfrac{1}{N}) + o (\delta_N^2)),\label{conv_order_MF}\allowdisplaybreaks\\
     %\mu_t^{i,\ast} = -R_k^{-1}[(B_k^\trns \Pi_k(t) + S_k^\trns) x_t^{i,\Phi} + (B_k^\trns s_k(t)+n_k)].
     &(c)~~\Big\vert J_0^{Q_j,N}(u_0,u_{-0}^{Q_j,\circ}) - J_0^{Q_j,\infty}(u_0) \Big\vert \nonumber\\&\hspace{3cm}\leq C(o(\tfrac{1}{\sqrt{N}}) + o(\delta_N)),\label{conv_order_cost}
\end{align}
\end{itemize}
where $x^{(N^{Q_j})}$, $x_0^{Q_j,N}$and $x_0^{Q_j}$ denote, respectively, the empirical state average, the major agent's statge in the finite-population case, and the major agent's state in the infinite-population case, at the discrete state $Q_j$. Moreover, $J_0^{Q_j,N}(.,.)$ and $J_0^{Q_j,\infty}(.)$ denote, respectively, the major agent's cost functional in the finite-population and the infinite-population cases at the discrete state $Q_j$. The dynamics governing the major agent in the finite-population and the infinite-population cases for an arbitrary control action $u_0 \in {\mc{U}}_{g}^{N,L}$ are, respectively,  given by
\begin{align}
&dx_0^{Q_j,N} = (A_0^{Q_j} x_0^{Q_j,N} + B_0^{Q_j} u_0 \nonumber\\&\hspace{2.5cm}+ F_0^{Q_j} x^{(N^{Q_j})} )dt + D_0^{Q_j} dw_0,\label{MajorStandardDynamics_hybrid-ApC} \\
&dx_0^{Q_j} = (A_0^{Q_j} x_0^{Q_j} + B_0^{Q_j} u_0 \nonumber\\&\hspace{2.1cm}+ \pi^{Q_j} \otimes F_0^{Q_j} \bar{x}^{Q_j} )dt + D_0^{Q_j} dw_0. \label{MajorStandardDynamics_hybrid_infty-ApC}
\end{align}
Following along the lines of the proofs of Theorem 6 and Proposition 8 in \cite{Huang2010}, we can show that \eqref{conv_order_state} holds for all $Q_j, 0\leq j \leq 3$.  

The difference between the major agent's cost functional in the finite-population and the infinite-population cases is given by 
\begin{multline} 
 J_0^{Q_j,N} - J_0^{Q_j,\infty}  = \hf\mathbb{E} \big [ \Vert x_0^{Q_j,N}(T) \Vert^2_{\bar{P}_0^{Q_j}}-\Vert x_0^{Q_j}(T) \Vert^2_{\bar{P}_0^{Q_j}} \\+ \int_{0}^{T}\Big(\Vert x_0^{Q_j,N}(s) - H_0^{Q_j} x^{(N^{Q_j})}(s)\Vert_{P_0^{Q_j}}^2 \\- \Vert x_0^{Q_j}(s) - H_0^{Q_j} \bar{x}^{Q_j}(s)\Vert_{P_0^{Q_j}}^2 \Big) ds \\
 = \mathbb{E} \bigg [ \Vert x_0^{Q_j,N}(T) - x_0^{Q_j}(T)\Vert^2_{\bar{P}_0^{Q_j}}+2[x_0^{Q_j}(T)]^T\bar{P}_0^{Q_j}\\ \times(x_0^{Q_j,N}(T) - x_0^{Q_j}(T))+\int_0^T\bigg(\Vert x_0^{Q_j,N}(s)-x_0^{Q_j}(s)\Vert_{P^{Q_j}_0}^2 \\+ 2 [x_0^{Q_j}(s)]^TP^{Q_j}_0(x_0^{Q_j,N}(s)-x_0^{Q_j}(s))
 \\+ \Vert x^{(N^{Q_j})}(s) - \bar{x}^{Q_j}(s)\Vert_{{(H_0^{Q_j})^TP_0^{Q_j}H_0^{Q_j}}}^2\\ + 2[H_0^{Q_j}\bar{x}^{Q_j}(s)]^TP_0^{Q_j}H_0^{Q_j}(x^{(N^{Q_j})(s)}-\bar{x}^{Q_j}(s))\\
 -2 (x_0^{Q_j,N})^T P_0^{Q_j}H_0^{Q_j}(x^{(N^{Q_j})}-\bar{x}^{Q_j}) \\-2(\bar{x}^{Q_j})^TP_0^{Q_j}H_0^{Q_j}(x_0^{Q_j, N}-x_0^{Q_j})\bigg)ds \bigg].\label{majorCostFunction_hybrid_Ap}
 %\Phi(x^{(N_t)}) \coloneqq H_0^k x^{(N_t)},
 \end{multline}
Using the Cauchy–Schwarz inequality, we can write 
\begin{multline}
    \mathbb{E} \big[(x_0^{Q_j})^TP^{Q_j}_0(x_0^{Q_j,N}-x_0^{Q_j})\big] \\\leq C\, \mathbb{E} \big[\Vert x_0^{Q_j}\Vert^2\big]^{\frac{1}{2}} \, \mathbb{E} \big[\Vert x_0^{Q_j,N}-x_0^{Q_j}\Vert^2\big]^{\frac{1}{2}}. 
\end{multline}
In a similar manner we can apply the Cauchy-Schwarz inequality to every cross term in \eqref{majorCostFunction_hybrid_Ap}. Hence from  \eqref{conv_order_state}-\eqref{conv_order_MF}, every cross term is at most of order $( o(\tfrac{1}{\sqrt{N}}) + o (\delta_N))$, and every squared term is of order $( o(\tfrac{1}{N}) + o (\delta_N^2))$. Therefore, we obtain \eqref{conv_order_cost}.
\begin{itemize}
\item[(ii)] From \eqref{conv_order_cost} we have
\begin{multline}\label{eps_ineq_arbitCntrl}
-o(\delta_N) - o(\tfrac{1}{\sqrt{N}}) 
 \leq J_0^{Q_j,N}(u_0,u_{-0}^{Q_j,\circ}) - J_0^{Q_j,\infty}(u_0) \\\leq o(\delta_N) + o(\tfrac{1}{\sqrt{N}}).
\end{multline}
Since \eqref{eps_ineq_arbitCntrl} holds for every $u_{0}$, from its right-hand side we have
\begin{equation}
J_0^{Q_j,N}(u_0^{Q_j,\circ},u_{-0}^{Q_j,\circ}) - o(\delta_N) - o(\tfrac{1}{\sqrt{N}}) \leq J_0^{Q_j,\infty}(u_0^{Q_j,\circ}),\label{eq1}
\end{equation}
and from its left-hand side we have 
\begin{gather}
-o(\delta_N) - o(\tfrac{1}{\sqrt{N}})
 \leq \inf_{u_0} J_0^{Q_j,N}(u_0,u_{-0}^{Q_j,\circ}) - \inf_{u_0} J_0^{Q_j,\infty}(u_0). \label{eq2}
\end{gather}
We then isolate $\inf_{u_0} J_0^{Q_j,N}(u_0,u_{-0}^{Q_j,\circ})$ on the right-hand side of \eqref{eq2} and use \eqref{eq1} to write
%\begin{align*}
%J_0^{Q_j,(N)}(u_0^{\circ},u_{-0}^\circ)& - 2o(\delta_N) - 2o(\tfrac{1}{\sqrt{N}}) \nonumber\allowdisplaybreaks\\&\hspace{0cm} \leq \inf_{u_0} J_0^{Q_j,\infty}(u_0) \leq \inf_{u_0} J_0^{Q_j,(N)}(u_0,u_{-0}^\circ).\quad
%\hfill \square
%\end{align*} 
\begin{align*}
J_0^{Q_j,N}(u_0^{Q_j,\circ},u_{-0}^{Q_j,\circ})& - 2o(\delta_N) - 2o(\tfrac{1}{\sqrt{N}}) \nonumber\allowdisplaybreaks\\&\hspace{0cm} \leq \inf_{u_0} J_0^{Q_j,N}(u_0,u_{-0}^{Q_j,\circ}).\quad
\hfill \square
\end{align*} 
\end{itemize}

\end{document}